\documentclass[twocolumn,showpacs,preprintnumbers,amsmath,amssymb]{revtex4} 

\usepackage{graphicx}
\usepackage{amsmath,amssymb}
\usepackage{epsfig}
\usepackage{dcolumn}
\usepackage{bm}
\usepackage{rotating}
\usepackage{times}
\usepackage{color}
\newcommand {\be}{\begin{equation}}
\newcommand {\ee}{\end{equation}}
\newcommand {\bea}{\begin{eqnarray}}
\newcommand {\eea}{\end{eqnarray}}
\newcommand {\bem}{\begin{displaymath}}
\newcommand {\eem}{\end{displaymath}}
\newcommand {\p}{\partial}
\newcommand {\f}{\frac }

\begin{document}

\preprint{ }

\title{Heat exchange mediated by a quantum system}

\author{George Y. Panasyuk}
\email{George.Panasyuk.ctr@wpafb.af.mil}
\author{ George A. Levin}
\author{ Kirk L. Yerkes}
\affiliation{Aerospace Systems Directorate,  Air Force Research Laboratory,  
Wright-Patterson Air Force Base, OH 45433}

\date{\today}

\begin{abstract}
We consider heat transfer between two thermal reservoirs mediated by a 
quantum system using the generalized quantum Langevin equation. 
The thermal reservoirs are treated as ensembles of  oscillators within the 
framework of the Drude-Ullersma model. General expressions for the heat current 
and thermal conductance are obtained for arbitrary coupling strength between 
the reservoirs and the mediator and for different temperature regimes. As 
an application of these results we discuss the origin of Fourier's law
in a chain of large, but finite subsystems coupled to each other by the quantum mediators.  
We also address a question of anomalously large heat current between the STM tip and substrate
found in a recent experiment. The question of minimum thermal conductivity is 
revisited in the framework of scaling theory as a potential application of the developed approach.

\end{abstract}


\maketitle

\section{INTRODUCTION}

The development of non-equilibrium thermodynamics mainly goes along two distinct directions.
One direction is the study of the energy transfer through  microscopic systems, such as nanotubes, molecules, 
or quantum dots~\cite{Dhar, Dubi_Di_Ventra}. Beyond a purely academic interest in the problem, 
research suggests that nanoscale 
and molecular systems are good candidates for many technological advances, such as molecular
wires, molecular diodes, rectifiers, and switches \cite{Jortner, Hanggi}.
The other direction, with longer history, deals with thermalization of, and energy flow through
finite, macroscopic subsystems. Examples of such an approach are the Caldeira-Legett \cite{Caldeira}
and Nieuwenhuizen-Allahverdian \cite{Allahverdian, Nieuwenhuizen} models in which a thermal reservoir 
or macroscopic system is described as a large or infinite ensemble of harmonic modes. 

In microscopic systems, such as chains of multilevel systems \cite{Michel1, dubi_diventra}, 
harmonic oscillators \cite{Michel2,Segal_Nitzan_Hanggi,Michel3}, or spins \cite{Saito2} the 
relaxation processes of individual elements and the processes of mutual equilibration between  
different elements are inseparable from each other and all take place on a microscopic time scale. 
In contrast, the local equilibrium requirement for macroscopic systems implies that the equilibration
processes  proceed on two vastly different time scales. The local equilibrium is 
established on the microscopic time scale, while the equilibration between the macroscopic subsystems  
takes a much longer time. 

One of the most visible problems of non-equilibrium thermodynamics is the microscopic derivation 
of Fourier's law,
specifying that the heat 
flux ${\bf j}$ through both fluids and solids is given by 
${\bf j} = -\kappa \nabla T({\bf r})$, where the temperature $T$ varies smoothly 
on microscopic scale and $\kappa$ is the thermal conductivity. Despite the 
ubiquitous occurrence of this phenomenon, very few rigorous mathematical derivations of 
this law are known~\cite{Boneto}. While for 3D generic models Fourier's 
law is expected to be true, this law may not be valid for 1D and 2D 
systems~\cite{Dhar}. 

A recently developed approach to study heat transport at the microscopic
level is based on either the classical or the quantum Langevin equation. 
The quantum Langevin equation
was first considered in Ref.~\cite{Senitzky} for a weakly damped harmonic oscillator.
Later~\cite{Mori}, it was used to formulate transport, collective motion, and the
Brownian motion from a unified, statistical-mechanical point of view. 
In Refs.~\cite{Ford, Caldeira, Haken, Klimontovich, Allahverdian, Nieuwenhuizen},
the Langevin equation was used for studying the thermalization of a 
particle coupled harmonically to a thermal reservoir and other closely-related problems.
This approach was generalized 
in Refs.~\cite{Zurcher, Saito, Dhar_Shastry, Segal_Nitzan_Hanggi} in order to explore
the non-equilibrium steady-state heat current and temperature profiles
in chains of harmonic oscillators placed between two thermal baths.
Recently, a new method for an exact solution to the Lindblad and Redfield master equations for
open quadratic system of $n$ fermions in terms of diagonalization of a $4n\times 4n$
matrix has been developed ~\cite{Prosen, Prosen1}. 
The method has been applied to Heisenberg $XY$ spin 1/2 chain coupled to heat baths at its ends.  
Generally, this approach can be considered as an  alternative to 
the  quantum Langevin equation. 

In this paper we investigate the non-equilibrium steady-state heat transfer between two thermal reservoirs described
as ensembles of harmonic modes mediated
by a quantum system, which is also considered in the harmonic approximation. 
This is a Hamiltonian system with fixed total energy, but increasing entropy. 
Our approach is based on the quantum Langevin equation and uses the Drude-Ullersma
model (DUM) for the bath mode spectrum. This is a generalization 
for the non-equilibrium situation of the approach employed in Ref.~\cite{Nieuwenhuizen} 
to the study of statistical thermodynamics of a quantum Brownian particle coupled  to a single thermal reservoir.

The solution obtained within this model allows us to determine 
the heat conductance between two thermal baths at arbitrary strength of the coupling constant. 
The results presented here are valid for an arbitrary temperature difference and arbitrary cut-off frequency,
which plays the role of the Debye frequency. 
As is found, temperature dependence of the conductance may possess a plateau
at intermediate temperature range, similar to the ``classical'' plateau at high 
temperatures. Dependence of the thermal conductance on the coupling strength 
displays a maximum.
We also show that the quantum thermal bath approach, in 
which a many-body
problem is replaced by the one-body approximation where the effect of thermal baths is quantified by 
random forces, yields the same
results for conductance as the rigorous (many-body) solution in the limit of large Debye frequency. 
On a more general note, this approximation can be successfully used for solving more complex problems
 without necessarily resorting on the rigorous solution that is based on the full-fledged Hamiltonian.

The solution to the problem of heat transfer between two thermal reservoirs with a quantum particle as the mediator
is applied to a chain system consisting of macroscopic subsystems coupled to each other by quantum particles (mediators). 
The microscopic time scale $\tau$ describes the time it takes for the heat current facilitated by the mediators
to come to a steady-state. 
Each subsystem has arbitrary large heat capacity and the equilibration time between them is much longer 
than $\tau$. In this case,  
Fourier's law follows naturally as the differential form of the energy conservation law. 

We use these results to explain a recent experiment in which the heat flow in vacuum 
between an STM tip 
and a substrate  was found to be about ten orders of magnitude greater than that expected from 
the blackbody radiation theory~\cite{Altfeder}. Our suggestion is that the heat flow in this experiment 
was mediated by a carbon monoxide molecule placed in the gap between the tip and substrate.
In addition, we briefly  discuss the problem of minimum thermal conductivity attained 
when the coherence length of the phonons is minimal and of the order of the interatomic distance.  
Finally, we also briefly mention a possible application of the developed model to 
study the Josephson junction, which provides an important example of strong coupling between 
the quantum system and thermal baths.

The paper is organized as follows. The  model is 
introduced in Sec. II, where the generalized Langevin equation is derived and solved
using the DUM. In Sec. III, expressions for the heat current between the thermal baths and
heat conductance are derived and analyzed for different temperature regimes and different 
coupling strengths. In Sec. IV, we compare the solution obtained in
the quantum thermal bath approach with the the rigorous solution.
In Section V the specifics of strong coupling is described. 
Section VI is devoted to Fourier's law in a chain of macroscopic subsystems.
Section VII discusses the  application of our model to anomalously large heat flow between 
the STM tip and substrate.
Sections VIII and IX discuss possible applications of the model to deal with the problem of 
minimum thermal conductivity and the Josephson junction, respectively.

\section{LANGEVIN EQUATION}

\begin{figure}[bottom]                                                     
\includegraphics[width=4.0cm]{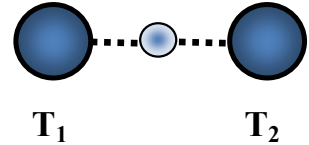}                                    
\caption{\label{ham}(Color online) Diagram representation of the total 
Hamiltonian (\ref{Htot}). The large circles correspond to the Hamiltonians of the thermal 
reservoirs, Eqs. (\ref{HBnu}),
the small circle is the Hamiltonian (\ref{H}) of the mediator, and the dotted lines correspond 
to the coupling Hamiltonians (\ref{Vnu}).}
\end{figure}

The total Hamiltonian of the system under consideration is similar to that in 
Refs.~\cite{Ford_Lewis_Connell, Segal_Nitzan_Hanggi}
\be
\label{Htot}
{\mathcal H}_{\rm tot} = {\mathcal H} + {\mathcal H}_{\rm B1} + {\mathcal H}_{\rm B2} + 
 {\mathcal V}_1 + {\mathcal V}_2 .
\ee
Here
\be
\label{H}
\mathcal H = \f{p^2}{2m} +  \f{k x^2}{2}
\ee
is the Hamiltonian of the quantum system (the mediator) described as a harmonic oscillator,
\be
\label{HBnu}
 \mathcal H_{{\rm B}\nu} = \sum_i \left [ \f{p_{\nu i}^2}{2m_{\nu i}} + 
\f{m_{\nu i}\omega_{\nu i}^2x_{\nu i}^2}{2}\right ] 
\ee
are the Hamiltonians of the first  ($\nu$ = 1) and second  ($\nu$ = 2) baths, and
\be
\label{Vnu}      
 \mathcal V_{\nu} =  -x\sum_i C_{\nu i}x_{\nu i} + 
x^2\sum_i \f{C_{\nu i}^2}{2m_{\nu i} \omega_{\nu i}^2}
\ee
describe interaction between the mediator and
the baths. 
In Eq. (\ref{H}), $x$ and $p$ are the 
coordinate and momentum operators and
$m$ and $k$ are the mass and spring constant of the particle. 
In Eqs. (\ref{HBnu}) and (\ref{Vnu}), $x_{\nu i}$ and $p_{\nu i}$ are the coordinate 
and momentum operators, whereas $m_{\nu i}$ and 
$\omega_{\nu i}$ are the masses and frequencies of the oscillators for the 
$i$th mode that belong to the $\nu$th bath. Finally, $C_{\nu i}$ are the 
coupling coefficients that describe the interaction between the quantum system and 
the baths. The last contributions to the right-hand side of (\ref{Vnu}) are 
self-interaction terms, which guarantee that 
${\mathcal H}_{\rm B\nu} + {\mathcal V}_{\nu}$ are positively defined operators.
Fig. \ref{ham} contains graphical representation of  the Hamiltonian (\ref{Htot}). 
Here the large circles represent
the Hamiltonians of the baths (\ref{HBnu}), the smaller central circle stands for the 
particle Hamiltonian (\ref{H}), and the dashed lines describe the interaction between 
the central particle and the baths (\ref{Vnu}).

The Heisenberg equations for the coordinate and momentum operators 
corresponding to each bath and for our quantum system read
\be
\label{heis_xnui}
\dot{x}_{\nu i} = \f{p_{\nu i}}{m_{\nu i}} ,
\ee
\be
\label{heis_pnui}
\dot{p}_{\nu i} = -m_{\nu i}\omega_{\nu i}^2x_{\nu i} + C_{\nu i}x ,
\ee
\be
\label{heis_x}
\dot{x} = \f{p}{m} ,
\ee
and
\be
\label{heis_p0}
\dot{p} = 
-k x + \sum_{i\nu}C_{\nu i}x_{\nu i} - x\sum_{i\nu}\f{C_{\nu i}^2}{m_{\nu i} \omega_{\nu i}^2} .
\ee
Considering Eqs. (\ref{heis_xnui}) and (\ref{heis_pnui}) as a system of inhomogeneous 
equations with known inhomogeneity, its solution can be written as
\begin{eqnarray}
\label{heis_xnui_sol}
x_{\nu i}(t) = x_{\nu i}(0)\cos (\omega_{\nu i}t) + 
\f{p_{\nu i}(0)}{m_{\nu i}\omega_{\nu i}}\sin (\omega_{\nu i}t) + \nonumber \\
 \f{C_{\nu i}}{m_{\nu i}\omega_{\nu i}}\int_0^tds\sin [\omega_{\nu i}(t-s)]x(s) 
\end{eqnarray}
and
\begin{eqnarray}
\label{heis_pnui_sol}
p_{\nu i}(t) = m_{\nu i}\dot{x}_{\nu i}(t) = -m_{\nu i}\omega_{\nu i}x_{\nu i}(0)
\sin (\omega_{\nu i}t) + \nonumber \\
p_{\nu i}(0)\cos (\omega_{\nu i}t) +
 C_{\nu i}\int_0^tds\cos [\omega_{\nu i}(t-s)]x(s) .
\end{eqnarray}
Substituting (\ref{heis_xnui_sol}) into (\ref{heis_p0}) and integrating by parts, 
one obtains the quantum Langevin equation:
\be
\label{langevin}
m\ddot{x} = -k x(t) + \eta (t) - \int_0^tds\gamma (t-s)\dot{x}(s) - \gamma(t)x(0) ,
\ee
where
\begin{eqnarray}
\label{eta}
\eta (t) = \sum_{i\nu}C_{\nu i}\left [x_{\nu i}(0)\cos (\omega_{\nu i}t) + 
\f{p_{\nu i}(0)}{m_{\nu i}\omega_{\nu i}}\sin (\omega_{\nu i}t)\right ] \,\,\,
\end{eqnarray}
is the noise that comes from the baths and
\begin{eqnarray}
\label{gamma}
\gamma (t) = \sum_{i\nu}\f{C_{\nu i}^2}{m_{\nu i}\omega_{\nu i}^2}\cos (\omega_{\nu i}t) 
\end{eqnarray}
is the friction kernel, which takes into account the interaction of the quantum 
particle with both thermal reservoirs. 

\subsection{Drude-Ullersma model}
\label{subsec:DUM}

At this point we have made no specific assumptions about the properties of the 
Hamiltonians that describe  the thermal reservoirs.
The microscopic structure of the thermal reservoirs does not affect the nature of 
the energy exchange between them.
Therefore, one can choose a specific, physically meaningful  model of the 
Hamiltonians $ \mathcal H_{{\rm B}1}$ and $ \mathcal H_{{\rm B}2}$  without 
sacrificing the  generality of the results.

Here we employ the Drude-Ullersma model 
(DUM)~\cite{Ullersma,  Weiss, Nieuwenhuizen}. The model assumes that in the absence of the
interaction with the quantum system, each bath consists of uniformly spaced 
modes and introduces the following $\omega$-dependence for the coupling coefficients:
\be
\label{DUM}
\omega_{\nu k} = k\Delta_{\nu}, \,\,\,\, C_{\nu i} = 
\sqrt{\f{2\gamma_{\nu}m_{\nu i}\omega_{\nu i}^2\Delta_{\nu} D_{\nu}^2}{\pi (\omega_{\nu i}^2 +  D_{\nu}^2)}} 
\ee
where $ k = 1, 2, 3, ... N_{\nu}$. In Eq. (\ref{DUM}), $\Delta_{\nu}$ are the mode 
spacing constants,  $D_{\nu}$  are the characteristic cutoff 
frequencies qualitatively similar to the Debye frequency,  and $\gamma_{\nu}$ are 
the coupling constants 
between a given reservoir  and the mediator~\cite{Nieuwenhuizen}. 
Hereafter we assume for simplicity that $D_1 = D_2 \equiv D$. 
In the final results we take the limit  $N_{\nu} \rightarrow \infty$ and $\Delta_{\nu} \rightarrow 0$.

 Substituting Eq. (\ref{DUM}) 
into Eq. (\ref{gamma}) and replacing summation over $i$ by integration,
one arrives at the following expression for $\gamma (t)$:
\be
\label{gamma_DUM}
\gamma (t) = \gamma D e^{-D|t|} ,
\ee
where $\gamma = \gamma_1 + \gamma_2$.
Using (\ref{DUM}) and  (\ref{gamma_DUM}),  Eq. (\ref{langevin}) can be solved by
Laplace transformation~\cite{Laplace}:
\be
\label{langevin_solved}
x(t) =\dot{g}(t)x(0) + \f{1}{m}g(t)p(0) + \f{1}{m}\int_0^tdsg(t-s)\eta(s) .
\ee
Here
\be
\label{g}
g(t) = L^{-1}\left [ \f{1}{z^2 + w_0^2 + z{\hat\gamma}(z)}\right ] \equiv L^{-1}[\tilde g(z)],
\ee
where $\tilde g(z)$ is the Laplace transform $L$ of $g(t)$, $L^{-1}$ is the inverse 
Laplace transform, and
\be
\label{gamma_hat}
{\hat\gamma}(z) = \f{1}{m}L[\gamma(t)] = \f{D\hat \gamma}{D + z}, \,\,\hat \gamma \equiv \f{\gamma}{m} .
\ee
After substituting (\ref{gamma_hat}) into (\ref{g}), $g(t)$ can be presented as
\be
\label{g_final}
g(t) = L^{-1}[\tilde g(z)] = 
\sum_{n=1}^3g_ne^{-\mu_n t},
\ee
where
\be
\label{g_z}
\tilde g(z) = \f{D + z}{(D+z)(z^2+\omega_0^2)+D\hat \gamma z} .
\ee
Here $g_n$ are defined by the last two relations and $\mu_n$ are 
the roots of equation
\be
\label{3polynoml}
(\mu - D)(\mu^2 + \omega_0^2) + \hat \gamma D\mu = 0,
\ee
where $\omega_0 = \sqrt{k/m}$. 
Statistical thermodynamics of a quantum particle coupled to 
a thermal bath was considered 
in~\cite{Nieuwenhuizen} in the limit of large $D$ when
\be
\label{large_D} 
D >> \omega_0, \,\,\,1/\tau_{\rm p} = \gamma/m \equiv \hat \gamma , \,\,\,
1/\tau_{\rm x} \equiv k/\gamma = \omega_0^2\tau_{\rm p}. 
\ee
In this case,  
(\ref{g_final}) and (\ref{3polynoml}) give
\be
\label{roots}
\mu_{1,2} \approx \f{1}{2\tau_{\rm p}}(1 \mp r), \,\,\, 
\mu_3 \approx D - 1/\tau_{\rm p} >> |\mu_{1,2}| 
\ee
and 
\be
\label{g12}
g_1 = -g_2 \approx \f{\tau_{\rm p}}{r} ,
\ee 
where  $r = \sqrt{1 - 4\tau_{\rm p}/\tau_{\rm x}}$. If 
$\tau_{\rm p}/\tau_{\rm x} = (\omega_0\tau_{\rm p})^2 \leq 1/4$, $\mu_{1,2}$ are real and 
if $\tau_p << \tau_x $, the quantities $\tau_{\rm p}$ and $\tau_{\rm x}$ can be 
interpreted as the characteristic relaxation times for the momentum and
coordinate, respectively. 
Otherwise, when 
$\omega_0\tau_{\rm p} > 1/2$, $\mu_{1,2} = 1/2\tau_{\rm p} \mp i/\tau_0$, where 
$\tau_0 = (\omega_0^2 - {\hat \gamma}^2/4)^{-1/2}$ determine the oscillation 
time while $\tau_p$ again determines the damping time. In what follows, 
 however, we consider a more general case when
$D$, $\omega_0$, and $\tau_{\rm p}^{-1}$  may be comparable. In this case, 
the roots $\mu_n$ and coefficients $g_n$, 
where $n = 1, 2, 3$, are determined as analytical solutions 
of (\ref{3polynoml}) and from relation (\ref{g_final}), respectively.

\section{HEAT CURRENT}
\label{heat_flux}

Using (\ref{heis_xnui}) and (\ref{heis_pnui}),  one can easily show that the 
rate of change of the energy of a given thermal reservoir is given by
\be
\label{q_balance}
\f{d}{dt}\sum_i\left \langle  \f{p_{\nu i}^2}{2m_{\nu i}} + 
\f{m_{\nu i}\omega_{\nu i}^2x_{\nu i}^2}{2} \right \rangle \equiv
-\langle {\mathcal P}_{\nu}\rangle ,
\ee 
where 
\be
\label{flux_nu}
{\mathcal P}_{\nu} = 
-\sum_i\f{C_{\nu i}}{2m_{\nu i}}\langle p_{\nu i} x + x p_{\nu i}\rangle 
\ee 
is the work the quantum system performs on the $\nu$th 
bath per unit of time (the  power dispersed in the $\nu$th bath)~\cite{Zurcher}. 
In the  steady-state regime  the power acquired by one reservoir is taken from 
the other, so that the steady-state heat current $J_{\rm th}$ can be presented
as 
$J_{\rm th} =  \langle {\mathcal P}_1\rangle$ = -$\langle {\mathcal P}_2\rangle$
or in the symmetrical form
\be
\label{q_flux}
J_{\rm th} = \f{1}{2}\langle {\mathcal P}_1 - {\mathcal P}_2\rangle .
\ee 

Using  solutions 
for $p_{\nu i}(t)$ and $x(t)$ from Eqs. (\ref{heis_pnui_sol}) and 
(\ref{langevin_solved}), 
and  omitting here the  transient processes that wash out over  the short time 
$\tau = {\max}(\tau_{\rm p}, \tau_{\rm x})$ 
we get
\begin{eqnarray}
\label{Jth_less_1}
\langle {\mathcal P}_{\nu}\rangle  
\approx -\f{1}{2m}\sum_{i=1}\f{C_{\nu i}}{m_{\nu i}}[\cos (\omega_{\nu i}t)j_{\nu}^{(a)} 
\nonumber \\
 - m_{\nu i}\omega_{\nu i}\sin (\omega_{\nu i}t)j_{\nu}^{(b)} + C_{\nu i}j_{\nu}^{(c)} ],
\end{eqnarray} 
where
\begin{eqnarray}
\label{Jth_less_a}
j_{\nu i}^{(a)} = \int_0^tdsg(t-s)\langle p_{\nu i}(0)\eta (s) + \eta (s) p_{\nu i}(0)\rangle ,
\,\,\,\,
\end{eqnarray} 
\begin{eqnarray}
\label{Jth_less_b}
j_{\nu i}^{(b)} = \int_0^tdsg(t-s)\langle x_{\nu i}(0)\eta (s) + \eta (s) x_{\nu i}(0)\rangle ,
\,\,\,\,
\end{eqnarray} 
and
\begin{eqnarray}
\label{Jth_less_c}
j_{\nu i}^{(c)} = 
\langle \int_0^td\tau \cos \omega_{\nu i}(t-\tau)x(\tau) \int_0^tdsg(t-s)\eta (s) +
\nonumber \\ \int_0^tdsg(t-s)\eta (s) \int_0^td\tau \cos \omega_{\nu i}(t-\tau)x(\tau)\rangle .
\,\,\,\,\,\,
\end{eqnarray}

The derivation of the steady-state expressions for $j_1^{(a)}$, $j_1^{(b)}$, $j_1^{(c)}$ 
and ultimately for $J_{\rm th}$ depends on how the 
contact between the baths  is established. A physically meaningful model 
should yield the  same result regardless of how the coupling initially takes place.
To verify this we have considered two options. In one case, the
quantum system is attached simultaneously to both baths at time $t = 0$.
In the second  case, the quantum particle is coupled  first
to the first bath, reaches thermal equilibrium with it, and at a later 
moment (which is again $t$ = 0) it is coupled to the second bath. 

We have established that the steady-state heat current is the same in either scenario.
Below we give a derivation in the case of simultaneous coupling of the mediator to two 
thermal reservoirs at $t=0$.
We can assume that for $t \leq 0$ the
dynamic variables of the baths are determined by the usual 
expressions:
\begin{eqnarray}
\label{1st_way_x}
x_{\nu i}(t) = \sqrt{\f{\hbar}{2m_{\nu i}\omega_{\nu i}}}
(a_{\nu i}^+e^{i\omega_{\nu i}t} + a_{\nu i}e^{-i\omega_{\nu i}t})
\end{eqnarray}    
and 
\be
\label{1st_way_p}
p_{\nu i}(t) = m_{\nu i}\dot{x}_{\nu i}(t).
\ee
Here the creation and annihilation operators $a_{\nu i}$ and 
$a_{\nu i}^+$ satisfy 
$[a_{\nu i},a_{\nu^{\prime} k}^+] = \delta_{ik}\delta_{\nu \nu^{\prime}}$. The operators'
Gibbsian ensemble averages are determined by
\be
\label{1st_way_T}
\langle a_{\nu i}^+a_{\nu^{\prime} k} + a_{\nu^{\prime} k}a_{\nu i}^+\rangle  = 
\delta_{ik}\delta_{\nu \nu^{\prime}}\coth \left ( \f{\beta_{\nu}\hbar\omega_{\nu i}}{2}\right ),
\ee  
where $\beta_{\nu} = (k_{\rm B}T_{\nu})^{-1}$, which also result in
\begin{eqnarray}
\label{xx_pp}
\langle p_{\nu i}(0)p_{\nu^{\prime} j}(0)\rangle  = 
m_{\nu i}^2\omega_{\nu i}^2\langle x_{\nu i}(0)x_{\nu^\prime j}(0)\rangle = 
\nonumber
\\
\f{\hbar m_{\nu i}\omega_{\nu i}}{2}\delta_{ij}\delta_{\nu \nu^{\prime}}
\coth \left ( \f{\beta_{\nu}\hbar\omega_{\nu i}}{2}\right )
\end{eqnarray}
and 
\begin{eqnarray}
\label{xp}
\langle p_{\nu i}(0)x_{\nu^{\prime} j}(0) + x_{\nu^{\prime} j}(0)p_{\nu i}(0)\rangle  = 0 .
\end{eqnarray}
Using these equations,  as well as Eq. (\ref{eta}), the ensemble averages
$\langle p_{\nu i}(0)\eta (t) + \eta (t)p_{\nu i}(0)\rangle$, 
$\langle x_{\nu i}(0)\eta (t) + \eta (t)x_{\nu i}(0)\rangle$ can be found 
and $\langle {\mathcal P}_{\nu}\rangle$ can be written as
\be
\label{P_ab_c}
\langle {\mathcal P}_{\nu}\rangle = 
\langle {\mathcal P}_{\nu}\rangle^{(1)} + \langle {\mathcal P}_{\nu}\rangle^{(2)}\,\,\,\,\,
\ee
where
\begin{eqnarray}
\label{P_ab}
\langle {\mathcal P}_{\nu}\rangle^{(1)} = -\f{\hbar}{2m}\sum_{i=1}\f{C_{\nu i}^2}{m_{\nu i}}
\coth \left ( \f{\beta_{\nu}\hbar\omega_{\nu i}}{2}\right )\times
\nonumber
\\
\left [ \cos (\omega_{\nu i}t)\int_0^tdsg(t-s)\sin (\omega_{\nu i}s) - \right.
\nonumber
\\
\left. \sin (\omega_{\nu i}t\int_0^tdsg(t-s)\cos (\omega_{\nu i}s))\right]
\end{eqnarray}    
and
\be
\label{P_c}
\langle {\mathcal P}_{\nu}\rangle^{(2)} = 
-\f{\hbar}{2m}\sum_{i=1}\f{C_{\nu i}^2}{m_{\nu i}}j_{\nu i}^{(c)} .
\ee
Evaluating the integrals in (\ref{P_ab}) using (\ref{g_final}) and omitting exponentially 
decaying contributions, one can find
that all time-varying  terms, such as  those proportional to 
$\sin (\omega_{\nu i}t)\cos (\omega_{\nu i}t)$, etc.  cancel 
each other out.  It means that the steady-state heat current is truly time-independent 
and does not contain any fluctuating contributions. 
Using Eq. (\ref{DUM}) and substituting integration for summation we obtain
\be
\label{P_ab_final}
\langle {\mathcal P}_{\nu}\rangle^{(1)} = -\f{\hbar\gamma_{\nu}D^2}{\pi m}
\sum_{n=1}^3g_n\mu_n^2
\int_0^{\infty}
\f{d\omega \omega \coth (\beta_{\nu}\hbar\omega /2)}{(\omega^2+D^2)(\omega^2+\mu_n^2)}
\ee 
Similarly, Eq. (\ref{P_c}) can be rewritten as follows
\begin{eqnarray}
\label{P_c_fin}
\langle {\mathcal P}_{\nu}\rangle^{(2)} = 
-\f{\hbar\gamma_{\nu}D^2}{2m}\langle \int_0^td\tau S(t-\tau)x(\tau) 
\int_0^tdsg(t-s)\eta (s)
\nonumber \\
+ \int_0^tdsg(t-s)\eta (s) \int_0^td\tau \cos S(t-\tau)x(\tau)\rangle,\nonumber
\end{eqnarray}
where
\be
\label{S}
S(t) = \sum_{i=1}\f{C_{\nu i}^2}{\gamma_{\nu}D^2m_{\nu i}}\cos (\omega_{\nu i}t) = 
2\delta(t) - De^{-Dt} .
\ee
Using (\ref{eta}), (\ref{xx_pp}), and (\ref{xp}), 
$\langle {\mathcal P}_{\nu}\rangle^{(2)}$ can be obtained in a similar way, 
reaching its steady-state value when $t \rightarrow \infty$, and the heat current is given by
\begin{eqnarray}
\label{Jth_final}
J_{\rm th} = -\f{\hbar D^2}{2\pi \tau_{\rm p}}\sum_{n=1}^3g_n\mu_n^2\int_0^{\infty}\f{d\omega \omega 
[n_1(\omega)-n_2(\omega)]}{(D^2+\omega^2)(\mu_n^2+\omega^2)}, \,\,\,\,\,\,
\end{eqnarray}
where $n_{\nu}(\omega) = 1/[\exp (\hbar\omega \beta_{\nu}) - 1]$ are the 
phonon occupation number for the respective thermal reservoir. 
Here we assume for simplicity that $\gamma_{1}=\gamma_{2}$. 
In a more general case 
when  $\gamma_{1} \ne \gamma_{2}$, $J_{\rm th}$ will be determined by the same
expression (\ref{Jth_final}) by substituting 
$1/2\tau_{\rm p} \rightarrow 2\gamma_1 \gamma_2/(\gamma m)$.
It can be shown explicitly that the same result for $\langle {\mathcal P}_{\nu}\rangle^{(2)}$
and, eventually, for $J_{\rm th}$ can be obtained if one performs $\tau$-integration in 
$j_{\nu i}^{(c)}$ first and, finally, calculates the $i$-sum in (\ref{P_c}). This provides an
additional verification of the formula (\ref{S}) and expression (\ref{Jth_final}).
If $|T_1 - T_2| \ll (T_1 + T_2)/2 \equiv T$, Eq. (\ref{Jth_final}) determines 
the heat conductance $K$:
\begin{eqnarray}
\label{conductance}
K = -\lim_{\Delta T \rightarrow 0}\f{J_{\rm th}}{\Delta T} = 
-\f{\tau_{\rm h}^2k_{\rm B}D^2}{8\pi \tau_{\rm p}}\times \nonumber \\
\sum_{n=1}^3g_n\mu_n^2\int_0^{\infty}
\f{d\omega \omega^2{\rm cosech}^2(\beta\hbar\omega/2)}{(D^2+\omega^2)(\mu_n^2+\omega^2)}, 
\,\,\,\,\,\,
\end{eqnarray}
where $\Delta T = T_2 - T_1$ and $\tau_{\rm h} = \hbar/k_{\rm B}T$.

In the second scenario of consecutive coupling of the mediator to the thermal baths, 
one can consider initially  a closed 
system that describes the equilibrium (Gibbsian) state of the first thermal
bath plus the mediator.
The corresponding set of eigenvalues and eigenmodes  $\{ \nu_k , e_k \}$, 
can be determined by  diagonalization of the Hamiltonian 
${\mathcal H} + {\mathcal H}_{\rm B1} + {\mathcal V}_1 $. 
As was found in Ref.~\cite{Nieuwenhuizen}, the frequencies $\omega_{1k}$ of the 
unperturbed  modes  of the Hamiltonian 
${\mathcal H}_{\rm B1}$ 
are shifted due to the interaction with the mediator to the values
\be
\label{eigen_nu}
\nu_{1k} = \omega_{1k} - \f{\Delta_1}{\pi}\phi (\omega_k),
\ee
where $\phi (\omega)$ is a certain known function of the  parameters of the Hamiltonian.
In the limit of small  $\gamma_1$,  $\phi (\omega) \sim \gamma_1$.

In the Appendix we have shown that after coupling of this thermalized combined system 
to the second thermal bath with a 
different temperature, 
the same steady-state heat  current (\ref{Jth_final}) will be 
established despite the small difference in the microscopic makeup of the two  baths.
This  is important not only from the physical point of view that the 
steady-state energy current between two thermal reservoirs should not depend on the 
initial conditions,
but also for the derivation of Fourier's law as shown below.

It should be noted that the existence of a unique steady-state independent of the 
initial conditions 
cannot be taken for granted. There is substantial literature devoted to this 
problem in classical and quantum systems~\cite{Casher, Rubin, Benguria, Dhar_Wagh, Dhar_Sen, Dhar}. 
Ref. ~\cite{Dhar_Sen} considers the existence of the steady-state
and thermal equilibration in a system that represents a quantum wire coupled to two baths. 
It appears that a necessary, but not sufficient condition of uniqueness of the steady-state  is 
absence of the bound state in the spectrum of the thermal bath. 
In the model we are using here the bound state would manifest itself as an imaginary root
in Eq. (\ref{3polynoml}). This would result in  non-decaying oscillatory 
contribution to the steady-state energy current and its dependence on initial conditions.
However, the absence of the bound states does not solely guarantee  the uniqueness of the
steady-state or thermal equilibration~\cite{Dhar_Wagh, Dhar_Sen, Dhar}.
In this regard, our results  demonstrate explicitly the existence and uniqueness of the 
steady-state in the considered model.

\subsection{Different temperature regimes}
\label{different_T-regimes}

Expressions for the heat current and heat conductance can be simplified in the limits 
of high and low  
temperatures when $\hbar|\mu_n|/k_{\rm B}T_{\nu}\ll 1$ and $\hbar|\mu_n|/k_{\rm B}T_{\nu}\gg 1$, respectively.
Here $\mu_n$  ($n = 1, 2, 3$) are the roots of Eq. (\ref{3polynoml}).

In the high temperature limit, Eqs. (\ref{Jth_final}) and 
(\ref{conductance}) reduce to 
\be
\label{class_gen}
J_{\rm th} = -K(T_2 - T_1); \,\,\,\,
K \approx -\f{D k_{\rm B}}{4\tau_{\rm p}}\sum_{n=1}^3\f{g_n\mu_n}{D+\mu_n} .
\ee
The above sum can be written as
\be
\label{sum_class}
\sum_{n=1}^3\f{g_n\mu_n}{D+\mu_n} = -L[\dot g(t)]|_{z=D} = -D{\tilde g}(D)
\ee 
when using the well known properties of the Laplace transform and also relation
$\sum_{n=1}^3g_n\mu_n^k = [(-1)^k - 1]/2$, where $k$ = 0, 1, 2.
Thus, $K$ in (\ref{class_gen}) can be written as
\be
\label{K_class_gen}
K \approx \f{k_{\rm B}}{4\tau_{\rm p}}\f{2D^2}{[2(D^2+\omega_0^2)+\hat \gamma D]}.
\ee
In the limit (\ref{large_D}) of large $D$ we obtain
\be
\label{class}
K \approx \f{k_{\rm B}}{4\tau_{\rm p}} = \f{k_{\rm B}\hat \gamma}{4} .
\ee 

In the deep quantum regime (low temperatures), we find
\be
\label{quant_gen}
J_{\rm th} \approx \f{\pi^3 k_{\rm B}^4(T_1^4 - T_2^4)}{30\hbar^3\tau_{\rm p}}
\sum_{n=1}^3\f{g_n}{\mu_n^2} .
\ee
Using again the Laplace transform and (\ref{g_z}), we find
\be
\label{sum_quant}
\sum_{n=1}^3\f{g_n}{\mu_n^2} = -\f{d{\tilde g}(z=0)}{dz} = \f{\hat \gamma}{\omega_0^4} 
\ee
and finally have, in the quantum regime,
\be
\label{quant}
J_{\rm th} \approx \f{\pi^3 k_{\rm B}^4}{30\hbar^3\omega_0^4\tau_{\rm p}^2}(T_1^4 - T_2^4)
\,\,\,\,{\rm and}\,\,\,\,K \approx \f{2\pi^3 k_{\rm B}^4T^3}{15\hbar^3\omega_0^4\tau_{\rm p}^2}.
\ee 
The temperature dependence of $J_{\rm th}$ is the same as in the Stefan-Boltzmann law
\be
\label{SB}
J_{\rm SB} = A\f{\pi^2 k_{\rm B}^4}{60\hbar^3c^2}(T_1^4 - T_2^4) ,
\ee 
where $A$ is the area of a black body from which radiation emits. We will use
this observation in Sec. \ref{experiment} for discussion of experimental results found
in~\cite{Altfeder}.

\section{Quantum thermal baths}

\label{subsec:comparison}

\label{subsubsec:SNH}

Let us show now that the result for the heat current given by Eq. (\ref{Jth_final}) in the limit
$D\rightarrow\infty$ can be 
reproduced within the quantum thermal bath (QTB) approach. 
In the phenomenological QTB model the many-body problem described by the Hamiltonian (\ref{Htot}) is 
replaced by a one body problem in which  friction is introduced ``by hand'',
instead of being a logical consequence of energy redistribution between the practically infinite 
number of modes.
The other effect  of  a thermal bath  is modeled by a random force. 
The equation of motion for the mediator takes form of the Langevin equation
\be
\label{simple_mod}
m\ddot x + \gamma \dot x + m\omega_0^2 x = F_1(t) + F_2(t) ,
\ee 
where the stochastic forces (colored noise) $F_1$ and $F_2$ describe the effects of the two heat baths
with temperatures $T_1$ and $T_2$, respectively. We take  
$\langle F_{\nu}(t)\rangle$ = 0 and 
$\langle F_{\nu}(t) F_{\nu^{\prime}}(t^{\prime})\rangle = 
\delta_{\nu \nu^{\prime}}K_{\nu}(t - t^{\prime})$, where $\nu, \nu^{\prime} = 1, 2$ and 
$K_{\nu}(t)$ is determined by its Fourier transform as
\be
\label{K_Fourier}
K_{\nu}(t) = 
\f{1}{2\pi}\int_{-\infty}^{\infty}d\omega \tilde K_{\nu}(\omega)e^{-i\omega t}
\ee 
with 
\be
\label{quantum_corr}
\tilde K_{\nu}(\omega) = 
\f{\gamma}{2}\hbar \omega \coth (\hbar\omega/2k_{\rm B}T_{\nu}). 
\ee
The  correlators of the Fourier transforms of the random forces are given by
\be
\label{correlators}
\langle \tilde F_{\nu}(\omega) \tilde F_{\nu^{\prime}}(\omega^{\prime})\rangle = 
2\pi \delta(\omega + \omega^{\prime})\delta_{\nu \nu^{\prime}}\tilde K_{\nu}(\omega).
\ee
The solution of Eq. (\ref{simple_mod}) is given by
\be
\label{x_Fourier}
\tilde x(\omega) = 
-\f{1}{m}\f{\tilde F_1(\omega)+\tilde F_2(\omega)}{\omega^2 + i\hat\gamma\omega - \omega_0^2}, 
\,\,\, \hat\gamma = \gamma/m.
\ee 
The energy conservation law for the mediator is given by
\begin{eqnarray}
\label{E_central}
\f{d}{dt}\left \langle \f{m{\dot x}^2}{2} + \f{m\omega_0^2x^2}{2}\right \rangle = 
-\gamma \langle {\dot x}^2\rangle + \langle F_1{\dot x}\rangle + 
\langle F_2{\dot x}\rangle  ,\;\;
\end{eqnarray}
which in the steady-state corresponds to
$\gamma \langle {\dot x}^2\rangle = \langle F_1{\dot x}\rangle + \langle F_2{\dot x}\rangle$. 

We define heat current $J_{\rm th}^{(\rm s)}$ 
as the energy transferred per unit of time from the first to the second bath: 
$J_{\rm th}^{(\rm s)} = -dE_1/dt$, or 
\begin{eqnarray}
\label{simpl_heat_flux}
J_{\rm th}^{(\rm s)} = 
-\f{1}{2}\gamma \langle {\dot x}^2\rangle + \langle F_1{\dot x}\rangle = 
\f{1}{2}(\langle F_1{\dot x}\rangle - \langle F_2{\dot x}\rangle) .
\end{eqnarray} 
Here  $E_1$ is the internal energy of the first bath and we take that the energy dissipated 
by the mediator is equally split between the two baths. In general, the energy dissipated by 
the mediator can be  split between the thermal baths in arbitrary proportion, but then the 
correlators (\ref{E_central}) also will be proportional to the 
fraction of energy dissipated by the mediator in a given bath. 

Using Eqs. (\ref{x_Fourier}) and (\ref{correlators}),
we obtain
\begin{eqnarray}
\label{simpl_heat_flux_final}
J_{\rm th}^{(\rm s)} = 
\f{i}{4\pi m}\int_{-\infty}^{\infty}
\f{d\omega \omega [\tilde K_1(\omega)-\tilde K_2(\omega)]}{\omega^2 + i \hat\gamma\omega - \omega_0^2} . 
\end{eqnarray}
Taking into account expression (\ref{quantum_corr}), the heat current 
(\ref{simpl_heat_flux_final}) can be rewritten as
\begin{eqnarray}
\label{simpl_heat_flux_quantum_1}
J_{\rm th}^{(\rm s)} = 
\f{\hbar}{2\pi \tau_{\rm p}^2}\int_{0}^{\infty}
\f{d\omega \omega^3 [n_1(\omega)-n_2(\omega)]}{(\omega^2 - \omega_0^2)^2 + {\hat \gamma}^2{\omega}^2}
\end{eqnarray}
and the corresponding conductance is
\begin{eqnarray}
\label{simpl_conductance_quantum_1}
K^{(\rm s)} = 
\f{\tau_{\rm h}^2k_{\rm B}}{8\pi \tau_{\rm p}^2}\int_{0}^{\infty}
\f{d\omega \omega^4 {\rm cosech}^2(\beta\hbar\omega/2)}{(\omega^2 - \omega_0^2)^2 + {\hat \gamma}^2{\omega}^2}.
\end{eqnarray}

Thus, the heat current given by Eq. (\ref{simpl_heat_flux_quantum_1}) and obtained  from 
the phenomenological quantum thermal bath approach  coincides in the limit $D \rightarrow \infty$ with that 
given by Eq. (\ref{Jth_final}), which is the result of  rigorous solution of a microscopic 
many-body model of the thermal reservoirs.
This is a solid indication that the quantum thermal baths approach
can be used to address more complicated problems of energy transfer via quantum mediators 
without resorting to a full-blown treatment based on a many-body Hamiltonian such as the one given by Eq. (\ref{Htot}). 
It should be noted that the QTB model has been recently 
successfully used 
by Dammak et al.~\cite{Dammak}  for sampling quantum fluctuations within the framework of  molecular 
dynamics (MD) simulations. Using the QTB model, the authors reproduced several experimental data at low 
temperatures in a regime where quantum statistical effects cannot be neglected.
Our result here suggests that the MD approach can account for quantum statistical effects in
non-equilibrium situations as well.

Also we should mention that a powerful method for solving  problems involving open quadratic systems for 
fermions has been developed by Prosen \cite{Prosen} and Prosen and \u{Z}uncovi\u{c} \cite{Prosen1}. It would be 
interesting to see how a problem of non-equilibrium bosonic systems, like the one considered here, can be 
reformulated in terms of this novel approach of "third quantization".

\section{Weak and strong coupling regimes}
\label{subsec:coupling}

In this section we return to the analysis of our results emphasizing the effects of the weak and 
strong coupling on heat transfer as well as some interesting features in the behavior 
of the heat conductance.  
The main purpose of this section is to clarify the conditions which may allow us to assign  a certain 
temperature to the mediator. 
This is possible when the mediator is in a state of weak non-equilibrium. For a simple system with two 
degrees of freedom 
the weak non-equilibrium means that the  virial theorem is approximately satisfied. This condition 
depends on the coupling strength
$\gamma $ and temperature. Let us consider  the average potential and kinetic energy: 
\be
\label{Txp}
k_{\rm B}T_{\rm x} = \langle k x^2\rangle \,\,\,\,{\rm and}\,\,\,\, k_{\rm B}T_{\rm p} = \langle p^2/m\rangle . 
\ee
The imbalance between them  has been proven to be useful in determining the statistics of
a quantum particle coupled to a single heat bath~\cite{Nieuwenhuizen}. 
Even when the total system -- (heat bath + mediator) -- is in thermal equilibrium, the virial theorem 
is not satisfied for the mediator, namely $T_{\rm p}(T) - T_{\rm x}(T)$  is non-zero for any finite $\gamma $. 
This is a manifestation of quantum entanglement
between a particle and a single thermal bath~\cite{Nieuwenhuizen}. 
The difference between the average kinetic and potential energy of the mediator can  serve as a criterion 
for the coupling strength also in the non-equilibrium case, as shown below. 

Eqs. (\ref{eta}), (\ref{langevin_solved}), and (\ref{1st_way_T}), and the
relation $p = m\dot{x}$, one can obtain
\be
\label{Tx}
T_{\rm x} = \f{\hbar \omega_0^2\tau_{\rm p}S}{\pi k_{\rm B}}, \,\, 
\tau_{\rm p}S = \sum_{n=1}^3g_n[I_n(1 - \hat \mu_n^2) + I_{n1} + I_{n2}]
\ee 
and
\be
\label{Tp}
T_{\rm p} = -\f{\hbar}{\pi k_{\rm B}}\sum_{n=1}^3g_n\mu_n^2[I_n(1 - \hat \mu_n^2) + 
I_{n1} + I_{n2}] .
\ee 
Here $\hat \mu_n = \mu_n/D$,
\begin{eqnarray}
\label{In}
I_n = \int_0^\infty\f{x dx}{(x^2+1)(x^2+\hat \mu_n^2)} =  \nonumber \\
 \f{i [\rm {arctan} (a) - (\pi/2) \rm {sign}(b)] - \ln (|\hat \mu_n|^2)}{2(1 - \hat \mu_n^2)}
\end{eqnarray} 
with $a = (\mu_{n{\rm r}}^2 -  \mu_{n{\rm i}}^2)/(\mu_{n{\rm r}}\mu_{n{\rm i}})$, 
$b = \mu_{n{\rm r}}\mu_{n{\rm i}}$, $\mu_{n{\rm r}} = \rm {Re}(\mu_n)$, and
$\mu_{n{\rm i}} = \rm {Im}(\mu_n)$. Finally,
\begin{eqnarray}
\label{Innu}
I_{n\nu} = \int_0^\infty\f{x dx}{(e^x - 1)[x^2+ (\tau_{h\nu}\mu_n)^2]}
\end{eqnarray} 
where $\tau_{{\rm h}1,2} = \hbar/k_{\rm p}T_{1,2}$.

The weak and strong coupling can be defined in terms of the effective bath-particle interaction 
strength $\hat \gamma_{\rm D}$, where
\be
\label{gamma_D} 
\hat \gamma_{\rm D} = D^2\hat \gamma (D^2 + \omega_0^2)^{-1}.
\ee
If $\hat \gamma_{\rm D}\ll \omega_0$, one can find from (\ref{g_final})-(\ref{3polynoml}) that 
\be
\label{mu_small_gamma}
\mu_{1,2} \approx \mp i \omega_0 + \hat \gamma_{\rm D}/2, \,\, 
\mu_3 = D - \hat \gamma_{\rm D}
\ee
and
\be
\label{g_small_gamma}
g_{1,2} \approx \mp \f{i}{2\omega_0} - \f{\hat \gamma_{\rm D}}{2(D^2+\omega_0^2)}, \,\, 
 g_3 = \f{\hat \gamma_{\rm D}}{D^2+\omega_0^2}
\ee
If $\hat \gamma_{\rm D}\ll \omega_0$, the mediator can be described as an oscillator with relatively 
small effective friction.
However, this condition by itself is not sufficient to guarantee the virial theorem. 
Only if temperatures of both thermal reservoirs are sufficiently high, i.e.
\be
\label{small_entangl}
k_{\rm_B}T_{1,2}/ \hbar \gg \hat \gamma_{\rm D},
\ee 
we get approximately equal steady-state values of $T_{\rm x,p}$:
\be
\label{Txp_quant_stat}
k_{\rm B}T_{\rm x} \approx k_{\rm B}T_{\rm p} \approx \f{1}{2}[U(T_1) + U(T_2)].
\ee
Here $U(T) = \hbar\omega_0/2 + \hbar\omega_0[{\rm exp}(\hbar\omega_0/k_{\rm B}T) - 1]^{-1}$ 
is the  average energy of a quantum oscillator in thermal equilibrium. The inequality (\ref{small_entangl})
can be considered as the usual condition of applicability of the Gibbsian
statistics, when  interaction energy between subsystems of a large closed
system is small with respect to the internal energies of the subsystems~\cite{Landau}.
Thus, as long as the virial theorem  is preserved,
one can assign  to the mediator a certain temperature $T$  on the basis of Eq. (\ref{Txp_quant_stat}):
$U(T) = (1/2)[U(T_1) + U(T_2)]$. 
In the high temperature limit  $k_{\rm B}T_{1,2}\gg \hbar\omega_0$, this leads to
\be 
\label{Txp_class}
T_{\rm x} \approx T_{\rm p} \approx T \approx (T_1 + T_2)/2,
\ee 
as is expected.
   
For the case of moderate or strong coupling (overdamped mediator),
$T_{\rm x,p}$ acquire  $\gamma$-dependence and the condition 
(\ref{Txp_quant_stat}) is not satisfied any more.
\begin{figure}                                                                  
\includegraphics[width=7.0cm]{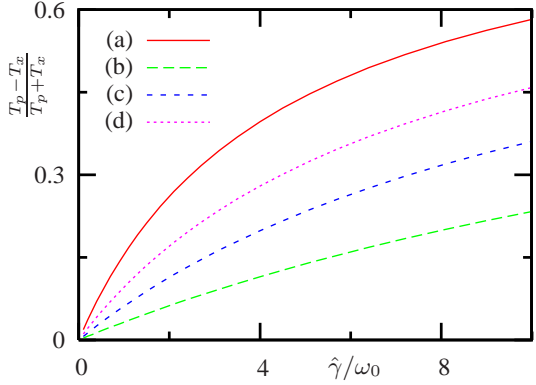}                               
\caption{\label{tem_diff}(Color online) Dependence of $\Delta \hat T_{\rm px}$ 
on the coupling strength for different $D/\omega_0$,
$\tau_{\rm h1}\omega_0$, and $\tau_{\rm h2}\omega_0$. (a) $D/\omega_0=10$, 
$\tau_{\rm h1}\omega_0=\tau_{\rm h2}\omega_0=1$, (b) $D/\omega_0=1$, $\tau_{\rm h1}\omega_0=\tau_{\rm h2}\omega_0=1$ 
(c) $D/\omega_0=10$, $\tau_{h1}\omega_0=\tau_{\rm h2}\omega_0=0.5$, 
and (d) $D/\omega_0=1$, $\tau_{\rm h1}\omega_0=1$, $\tau_{\rm h2}\omega_0=0.5$.}
\end{figure}
Fig. \ref{tem_diff} shows the monotonic dependence of the relative energy imbalance 
$\Delta \hat T_{\rm px} \equiv (T_{\rm p} - T_{\rm x})/(T_{\rm p} + T_{\rm x})$ 
on the coupling  strength.
The cut-off parameter $D$ can loosely associate with the Debye frequency.
The thermal bath modes with 
frequencies higher than $D$ are effectively decoupled from the mediator and do not play a significant 
role in the processes of thermalization and energy transfer. 
For this reason, when $D/\omega_0$ decreases, 
$\Delta \hat T_{\rm px}$ also decreases. 
This also follows from (\ref{gamma_D}). Indeed, the coupling constant $\hat \gamma \equiv \gamma/m$ 
is renormalized by the factor $D^2/(D^2 + \omega_0^2)$ and is effectively 
determined by $\hat \gamma_{\rm D}$ at a relatively small $\hat \gamma_{\rm D}$.
As one also finds, $\Delta \hat T_{\rm px}$ decreases when $T_{1,2}$ grow in accordance
to (\ref{small_entangl}) - (\ref{Txp_class}).
Thus, $\Delta \hat T_{\rm px}$ can be considered 
as a measure of the coupling strength that takes into account all the relevant factors 
such as $D^2/(D^2+\omega_0^2)$ and values of $T_{1,2}$.

\begin{figure}[here]                                                      
\includegraphics[width=7.0cm]{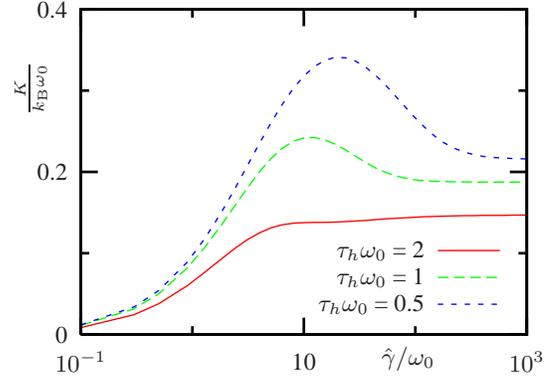}                     
\caption{\label{cond}(Color online) Dependence of the normalized heat 
conductance $K/(k_{\rm B}\omega_0)$ on the coupling strength at $D/\omega_0=1$ and
different $\tau_{\rm h}\omega_0$.}
\end{figure}

It is interesting to notice that the
$\gamma$-dependence of the heat conductance (\ref{conductance}) 
may possess a maximum. As we found, it appears at relatively small $\omega_0$ 
and relatively large $\hat \gamma$ (strong quantum entanglement between the particle and baths)
when $\tau_{\rm h}\omega_0 \lesssim 2$, $\hbar \hat \gamma/k_{\rm B}T \gtrsim 10$, and $\hat \gamma/D > 1$ as 
is illustrated in Fig. \ref{cond}.
The maximum strength can be characterized by the quantity 
$\Delta \hat K = (K_{\rm max} - K_{\infty})/K_{\rm max}$, where $K_{\rm max}$ and $K_{\infty}$ are
the values of $K$ at its maximum and at $\hat \gamma/\omega_0 \rightarrow \infty$, respectively.
As our simulations show, $\Delta \hat K$ increases when $\tau_{\rm h}\omega_0$ decreases and the 
value $(\hat \gamma/\omega_0)_{\rm max}$ at which $K_{\rm max}$ is achieved shifts toward larger values
when $D/\omega_0$ decreases.
It is worth to mention that a similar maximum in the dependence of the heat 
current on the system-bath coupling strength was found in~\cite{Prosen1} for 
Heisenberg $XY$ spin 1/2 chain.
\begin{figure}[here]                                                         
\center
\includegraphics[width=7.0cm]{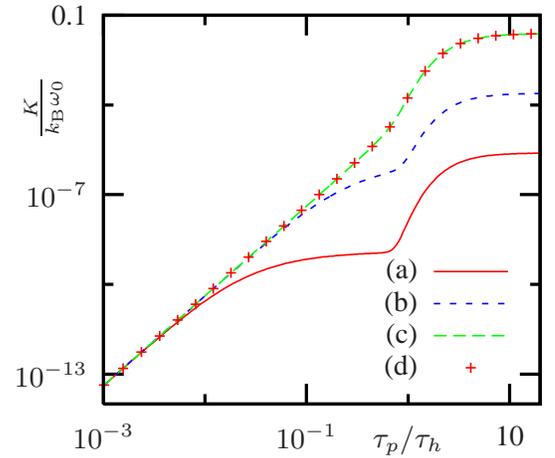}                      
\caption{\label{condTnew}(Color online) Temperature dependence for the normalized 
heat conductance $K/(k_{\rm B}\omega_0)$ at $\omega_0\tau_{\rm p} = 10$ and 
different values of $D$. Note that $\tau_{\rm p}/\tau_{\rm h}  \equiv k_{\rm B}T/\hbar\hat\gamma$.  
(a) $D\tau_{\rm p}$ = 0.1,  (b) $D\tau_{\rm p}$ = 1, 
(c) $D\tau_{\rm p}$ = 100, and (d) $D\tau_{\rm p} = \infty$.}
\end{figure}

Fig. \ref{condTnew} shows the temperature dependences for the normalized heat conductance at 
different $D$ and for
$\omega_0\tau_{\rm p} = 10$. The shown dependencies are
based on the same expression (\ref{conductance}) and its high-$D$ limit. 
As our analysis reveals,  if $D/\omega_0 \gtrsim 10$, $K(T)$ essentially coincides
with the corresponding $D \rightarrow \infty$ limit 
(\ref{simpl_conductance_quantum_1}). 
The straight line region corresponds to the low-$T$ limit (\ref{quant}),
which is the same for all curves. 
At large $T$, each curve reaches its classical plateau in accordance to 
(\ref{class_gen}).
An unusual feature, which is an additional 
plateau that appears in the intermediate range of $\tau_{\rm p}/\tau_{\rm h}$ and 
$D << \omega_0$, is a generic result valid in a wide range 
of $\omega_0\tau_{\rm p}$.
It can be explained in the following way. If $T$ is so large that 
$\omega_0\tau_{\rm h} < 1$ or $\tau_{\rm p}/\tau_{\rm h} > \omega_0\tau_{\rm p}$, we have the "final" 
classical plateau 
in Fig. \ref{condTnew} described by (\ref{K_class_gen}). On the other hand, if $T$ is so small 
that $D\tau_{\rm h} > 1$ or $\tau_{\rm p}/\tau_{\rm h} < D\tau_p$, we have the quantum regime (straight line in 
Fig. \ref{condTnew}) described by (\ref{quant}). In the intermediate region, when $D\tau_{\rm h} < 1$ 
but $\omega_0\tau_{\rm h} > 1$ or 
$D\tau_{\rm p} < \tau_{\rm p}/\tau_{\rm h} < \omega_0\tau_{\rm p}$, one can approximate 
$\omega^2+\mu_{1,2}^2 \approx \mu_{1,2}^2$ under the integrals in (\ref{conductance}) for $n$ = 1 or 2 and
\begin{eqnarray}
\label{conductance_plat}
K \approx 
-\f{\tau_{\rm h}^2k_{\rm B}D^2}{8\pi \tau_{\rm p}}\left [ J_1
\sum_{n=1,2}g_n + 
g_3D^2J_2\right ],
\end{eqnarray}
where
\be
\label{J_k}
J_k = \int_0^{\infty}\f{d\omega \omega^2}{{\rm sinh}^2(\beta\hbar\omega/2)(D^2+\omega^2)^k}.
\ee 
Taking into account that $g_1+g_2 = -g_3$, one can rewrite (\ref{conductance_plat}) as
\begin{eqnarray}
\label{conductance_plat_1}
K \approx 
\f{\tau_{\rm h}^2k_{\rm B}D^2}{8\pi \tau_{\rm p}}g_3 
\int_0^{\infty}\f{d\omega \omega^4}{{\rm sinh}^2(\beta\hbar\omega/2)(D^2+\omega^2)^2}.
\end{eqnarray}
Finally, using that $D\tau_{\rm h} < 1$ and approximating ${\rm sinh}(x) \approx x$, the integral in 
(\ref{conductance_plat_1}) can be estimated as $\pi/(D\tau_{\rm h}^2)$. Due to (\ref{g_small_gamma}), 
\begin{eqnarray}
\label{conductance_plat_final}
K \approx \f{k_{\rm B}D}{8\pi \tau_{\rm p}}g_3 = \f{k_{\rm B}\hat \gamma}{4}\f{D^3\hat \gamma}{2\omega_0^4} 
\end{eqnarray}
does not, indeed, depend on temperature forming the plateau in Fig. \ref{condTnew}.
The physical origin for the small $K$ and intermediate plateau can be explained as follows.
If $T$ increases above the Debye temperature $\theta_{\rm D} = D\hbar/k_{\rm B}$, i.e. 
$D\tau_{\rm h} = \theta_{\rm D}/T <$ 1, all the baths modes begin to be excited. At the same time, if $T$
is still less than $\hbar \omega_0/k_{\rm B}$ (or $\tau_{\rm h}\omega_0 >$ 1, which is always possible if 
$\omega_0 \gg D$), the quantum system cannot be excited and, hence, it cannot absorb energy from either bath 
and transfer it to the other bath. This will lead to a small value of $K$ in (\ref{conductance_plat_final}).
Moreover, because this situation stays unchanged when $T$ changes within the interval
$\theta_{\rm D} < T < \hbar \omega_0/k_{\rm B}$ 
(or $D\tau_{\rm p} < \tau_{\rm p}/\tau_{\rm h} < \omega_0\tau_{\rm p}$), $K$ must not depend on 
$T$ noticeably, which is in accordance to (\ref{conductance_plat_final}). 
Also, as one can notice, the effective 
bath-particle interaction strength $\hat\gamma_{\rm D}$ is small if $D \gg \omega_0$ even if 
$\hat \gamma$ itself is not small.
In this case, we have a very low
decay rate for the  $\mu_{1,2}$ modes and, correspondingly, very small
heat current and heat conductance, again in accordance to (\ref{conductance_plat_final}).
This result brings our 
situation close to the one with the bound states mentioned at the end of the first part of
Sec. \ref{heat_flux}. Indeed, when $\hat\gamma_{\rm D}$ = 0, we would have non-decaying
oscillatory contributions to the steady-state, depending on the initial conditions, and
zero heat current. However, even for a case with infinitely small $\hat\gamma_{\rm D}$, the unique 
but infinitely small steady-state current will be established after an infinitely large time
interval.

\section{Fourier's law}
\label{Fourier_law}

\begin{figure}[bottom]                                                     
\includegraphics[width=8.0cm]{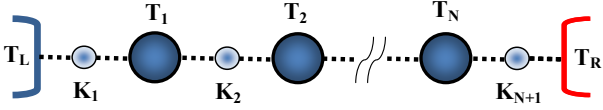}                                 
\caption{\label{chain}(Color online) Chain of macroscopic subsystems ("nanoparticles")  interconnected
by mediators. Each subsystem indicated by a large circle, as well as both thermal reservoirs,  
correspond to Hamiltonian
(\ref{HBnu}). Other symbols also have the same meaning as in Fig. \ref{ham}
Temperatures $T_{\rm L}$ and $T_{\rm R}$ of the left and right thermal 
reservoirs are fixed because their heat capacity is considered infinite.
The temperatures $T_n$ of the subsystems
can slowly vary until the steady-state is established. Thermal conductances
$K_n$ can vary from one connection to another.}
\end{figure} 
The results obtained in previous sections will be applied to an extended model 
that can shed additional light on the long-standing problem regarding the origin of Fourier's 
law~\cite{Boneto}.
The standard definition of a macroscopic body in the state of weak non-equilibrium \cite{Lifshitz} 
is that it can be divided into regions large enough to be considered macroscopic, but small 
enough to be described by a local temperature. In addition, these regions or subsystems  
must be weakly coupled to each other. The weak coupling in this context means that the 
characteristic time required for the subsystems to come to mutual  equilibrium is much longer 
than the time of microscopic relaxation. Here we introduce a model that fills this conceptual 
framework with a microscopic content.

Fig. \ref{chain}  illustrates our model. It consists of $N$ macroscopic subsystems and two 
thermal reservoirs (TR) coupled by the mediators. Each subsystem and each TR is described by 
the Hamiltonian given by Eq. (\ref{HBnu}) within the framework of the Drude-Ullersma model, 
Eq. (\ref{DUM}).  
Each  mediator is described by the Hamiltonian (\ref{H}) and each coupling is described by the 
Hamiltonian  (\ref{Vnu})  and Eq. (\ref{DUM}).
The total system, including the TRs, is Hamiltonian with constant total energy. 

Each subsystem and each TR  is initially prepared in the state of thermal equilibrium and 
is characterized by a temperature $T_n$ or $T_{\rm L}$ and $T_{\rm R}$, respectively. It means that the 
respective correlators have the form of Eq. (\ref{xx_pp}). 

The energy of a given subsystem, which consists of very large or infinite number of modes with frequencies 
$\omega_k=k\Delta $ with $k$ = 1, 2, 3 ... , is 
\be
U(T)\sim \f{(k_{\rm B} T)^2}{\hbar \Delta }\nonumber .
\ee
Here, the divergent zero-point energy term was dropped. 
Correspondingly, the heat capacity of a given subsystem
\be
C(T)=\f{d U}{d T}\sim k_{\rm B}\f{k_{\rm B} T}{\hbar \Delta }\nonumber .
\ee
Thus, the thermal reservoirs will be characterized by $\Delta\rightarrow 0$ and, correspondingly,  
an infinite heat capacity, while the subsystems need to be described by an arbitrarily small, 
but still finite $\Delta$. The energy difference between two subsystems (as in Fig. \ref{ham}) with 
temperatures $T_1$ and $T_2$  is 
\be
\Delta U\sim \f{|(k_{\rm B} T_1)^2-(k_{\rm B} T_2)^2 |}{\hbar \Delta }\sim 
\f{k_{\rm B}^2|T_1-T_2|T_{\rm av}}{\hbar \Delta }
\nonumber .
\ee
On the other hand, the heat current between the two subsystems can be estimated on the basis of 
Eqs. (\ref{class_gen}) and (\ref{class}) as
\be
J_{\rm th}\sim \f{k_{\rm B}}{\tau_{\rm p}}|T_1-T_2|.\nonumber
\ee
Thus, the characteristic time of mutual equilibration 
\be
\label{time}
t_{\rm eq} \sim \f{\Delta U}{J_{\rm th}} 
\sim \tau_{\rm p}\f{k_{\rm B}T_{\rm av}}{\hbar \Delta } = \f{\tau_{\rm p}}{\tau_{\rm h}}\Delta^{-1}
 \sim \Delta^{-1} .
\ee
For the subsystems, we assume that the ``Heisenberg'' time scale $1/\Delta$ is much larger
than any other characteristic times, such as $\tau_{\rm p}$, $\tau_{\rm x}$, 
or $\tau_{\rm h}$~\cite{Allahverdian,Nieuwenhuizen}.
Thus, the system described by the Hamiltonian represented by the 
diagram in Fig. \ref{chain} meets the conditions described in the first paragraph of this section. 

One of the main results that we have obtained by solving the Hamiltonian (\ref{Htot}) is that the 
energy flows from 
higher to lower temperature TR. Since all the modes of a given TR are in thermal equilibrium at 
the same temperature, the thermodynamic relationship between energy and entropy $dU_i=TdS_i$ is 
satisfied for each of them and the total entropy in the steady-state increases 
\be
\f{dS}{dt}=|J_{\rm th}|\left |\f{1}{T_1}-\f{1}{T_2}\right | >0. \nonumber
\ee
One can easily show that for the system represented in Fig. \ref{chain} the total entropy also increases even 
when the energy currents between the subsystems are all different. In the steady-state,
which corresponds to all energy currents between two neighboring subsystems being equal, the 
entropy increases as 
\be
\f{dS}{dt}=|J_{\rm th}|\left |\f{1}{T_{\rm L}}-\f{1}{T_{\rm R}}\right | >0. \nonumber
\ee

These results are valid as long as all modes remain in thermal equilibrium or close to equilibrium.
The coupling constants given by  Eq. (\ref{DUM}) are proportional to the infinitesimal parameter $\Delta$, 
\be
C_{\nu, i}\sim \Delta^{1/2}\nonumber.
\ee
The effect of such coupling on the mediator is finite because all modes contribute constructively. 
The rate of change of the correlators (\ref{xx_pp}) is much slower because each of the modes is coupled 
only to the mediator with vanishingly small coupling constant. One can see from 
Eqs. (\ref{heis_pnui}) - (\ref{eta}) that 
the rate of change of the correlators (\ref{xx_pp})
\be
\f{\p}{\p t}\langle p_{\nu i}(t)p_{\nu^{\prime} j}(t)\rangle  \sim
\f{\p}{\p t}\langle x_{\nu i}(t)x_{\nu^\prime j}(t)\rangle \sim \Delta. \nonumber
\ee
Thus, if we consider the evolution of the system on the time scale  $t$, such that
\be
\label{time scale}
\tau_{\rm p}<t \ll t_{\rm eq},
\ee
which is long enough for the microscopic relaxation to take place, but short on the macroscopic 
time scale, 
all modes will remain approximately in thermal equilibrium determined by the initial conditions.

The same argument allows us to extend the solution of the Hamiltonian (\ref{Htot}) shown as a diagram in 
Fig. \ref{ham} to the Hamiltonian that
corresponds to Fig. \ref{chain}.  Imagine that after we prepared both TRs and all the subsystems 
in the state of thermal equilibrium at the corresponding temperatures we start turning on 
the couplings to mediators  one-by-one from left to right.
The energy flow between two subsystems indexed as $n$ and $n+1$ will be the same as for the 
Hamiltonian (\ref{Htot}) because the effect of the subsystem $n$ being already coupled to the 
subsystem $n-1$  adds to the energy current a contribution of the  order of $\Delta $ and 
therefore negligible on the time scale  (\ref{time scale}). This means that the solutions 
of the Hamiltonian (\ref{Htot}) can be directly applied to the Hamiltonian of the chain shown in 
Fig. \ref{chain} in the form of energy conservation condition
\be
\label{E balance}
\partial_t U_n = J_{n-1,n} - J_{n,n+1} .
\ee 
Here the energy currents $J_{n-1,n}$ are given by Eq. (\ref{Jth_final}) and this equation is valid for 
arbitrary values of the initial temperatures.
In order to obtain Fourier's law we have to take the temperature differences between 
the neighboring subsystems small and express  the currents in terms of thermal 
conductances (\ref{conductance}) 
\be
\label{h_balance}
\partial_t U_n = K_{n-1,n}(T_{n-1} - T_n) - K_{n,n+1}(T_n - T_{n+1}) .
\ee   
This equation can be rewritten in the differential form by introducing a continuous 
coordinate $x$, where $x = n d$ corresponds to the locations of the subsystems and $d$ is the 
distance between them.
For identical mediators Eq. (\ref{h_balance}) can be recast as
\begin{eqnarray}
\label{h_balance1}
C(T)\partial_t T(x)= K(x-d/2)[T(x-d) - T(x)] - \nonumber \\
 K(x+d/2)[T(x) - T(x+d)],
\end{eqnarray}   
where $C(T) = dU/dT$ is the heat capacity of a subsystem. Then, Eq. (\ref{h_balance1}) 
leads to the energy conservation condition with Fourier's form of the energy current
\begin{eqnarray}
\label{Fourier}
\tilde {C}(T)\partial_t T(x)= \partial_x[\kappa(x)\partial_xT(x)],
\end{eqnarray}
where $\tilde {C}=C/d$ is the specific heat of the chain, and the thermal conductivity  $\kappa (T) = K(T)d$. 

This analysis shows that as long as the energy flow between macroscopic subsystems satisfies 
the condition of entropy increase, namely, that  the energy flows from higher  to lower 
temperature, Fourier's law is a straightforward consequence of energy conservation. 
What  has been proven in this paper is that the dynamics of the Hamiltonian system 
described by the Hamiltonian (\ref{Htot}) and its extension shown as a diagram in 
Fig. \ref{chain} does indeed lead to  the entropy increase. 
This statement is predicated on the condition of local thermal equilibrium or near 
equilibrium for the subsystems that exchange energy between themselves. We have stated  
in our treatment of the Hamiltonian (\ref{Htot}) that this is an initial condition. As such, 
it remains true for the time interval indicated by inequality (\ref{time scale}). However,  
the generally accepted understanding of the slow relaxation processes~\cite{Lifshitz} in 
a macroscopic body implies that as each  subsystem slowly gains or loses energy, it 
remains close to thermal equilibrium with a certain time-dependent temperature due to 
rapid thermalization on the microscopic time scale. Thus, the microscopic derivation 
of Fourier's law in the context of our model  requires not only to prove that in 
a certain limit (smooth temperature variation) the energy current is proportional to the 
temperature gradient, which we have done. 
Equal, if not more important, task is to show that the subsystems described by the 
Drude-Ullersma model are capable of self-thermalization when coupled by the harmonic 
mediator. This proof will require the study of the dynamics of the Hamiltonian (\ref{Htot}) on the 
time scale given by Eq. (\ref{time}). 

The model depicted in Fig. \ref{chain} resembles to some extent the models analyzed 
by Michel, Mahler and Gemmer \cite{Michel1}, as well as those discussed by
Dubi and Di Ventra \cite{dubi_diventra}. The subsystems considered 
in \cite{Michel1,dubi_diventra} were still microscopic with the heat capacity of the 
order of $k_{\rm B}$. Other publications \cite{Saito2, Michel2, Segal_Nitzan_Hanggi,Michel3} 
consider the energy transport in chains consisting of spins or harmonic oscillators.  
In all these cases the goal was to examine the energy transport on the "nano-scale". 
Our approach here is to examine a solvable Hamiltonian model of a non-equilibrium system 
that falls within the more traditional, "textbook" framework. The subsystems described 
by the Hamiltonian (\ref{HBnu}) can have arbitrarily large heat capacity determined by the 
infinitesimal parameter $\Delta$. Correspondingly, these subsystems remain in thermal 
equilibrium for the extended period of time, much greater than the microscopic relaxation time of 
the mediator and demonstrate that the 
energy flow leads to entropy increase and, hence, to Fourier's law, 
Eqs. (\ref{E balance}) - (\ref{Fourier}). The mediators do not have to be in the state of 
thermal equilibrium 
for the Eq. (\ref{Fourier}) to be valid. As was discussed in Sec. V,  in the strong 
coupling regime the mediator cannot be assigned a certain temperature. However, 
as long as the subsystems remain  close to  thermal equilibrium, Fourier's 
law is still valid. 

The outstanding question of the quasistatic evolution and self-thermalization of such 
subsystems on the much longer time scale, Eq. (\ref{time}),  
will be addressed elsewhere.

\subsection{Quantum thermal baths model}
 
\begin{figure}[bottom]                                                     
\includegraphics[width=8.0cm]{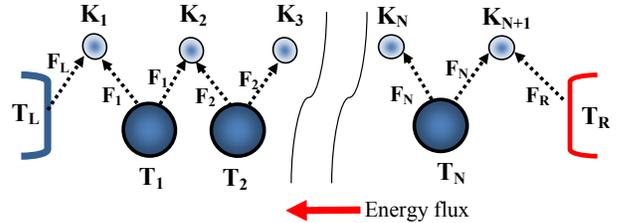}                              
\caption{\label{Fig5f}(Color online) Diagram of the
quantum thermal baths model equivalent to the Hamiltonian shown in Fig. \ref{chain}. Here $F_n$ are 
the random forces that appear
in Eq. (\ref{qb}) and the other elements are similar to ones from Fig. \ref{chain}.
Notice that the mediators are not coupled to each other directly.}
\end{figure}

In the limit of large $D$, 
Eq. (\ref{E balance}) follows from the model shown in Fig. \ref{Fig5f}. 
The equation of motion 
for the $n$'s mediator is practically the same as Eq. (\ref{simple_mod})
\be
\label{qb}
m\ddot x_n + \gamma \dot x_n + m\omega_0^2 x_n = F_{n-1}(t) + F_n(t) ,
\ee 
where the random force $F_n$ describes the effect of the corresponding subsystem $n$ 
on the mediator and the correlator of the force is determined by the temperature of 
the respective subsystem.  Note, that the mediators are not directly coupled to each 
other by "springs", so that Eqs. (\ref{qb}) are not a system of coupled equations, 
but simply are $n$ identical equations, whose solution for the heat current $J_{n-1,n}$ 
is given by  Eq. (\ref{simpl_heat_flux_quantum_1}). 

One can compare this model with the models of self-consistent 
reservoirs \cite{Bolsterli,Rich,Malay}. 
Since our subsystems are macroscopically large, their effect on the mediator can be 
described in terms of a stochastic force. 
However, these forces are not arbitrarily introduced. 
We have shown that the exact solution of the Hamiltonian dynamics in the limit of 
large $D$ yields the same expression for the energy current as the quantum thermal 
baths model. In this sense the quantum thermal baths model
can be a useful shortcut, but our  model is defined by the Hamiltonian (\ref{Htot}).  
It is also important to mention that the thermalization of these self-consistent 
reservoirs is considered a given in Refs. \cite{Bolsterli,Rich,Malay}. In fact, 
it has to be proven by analysis of the long term evolution of the Hamiltonian dynamics. 

One should note that in our model the mediators are not coupled to each other directly. 
The energy flows only through the macroscopic subsystems. In this sense the chain of 
mediators operate, by design, in the minimum thermal conductivity limit  because 
the neighboring mediators are always uncorrelated.  If we were to associate the movement 
of the mediators with phonons, their mean free path would be  minimum possible, equal 
to "interatomic" distance.

\section{Heat current between STM tip and substrate}

\label{experiment}

The results obtained in previous sections can be used to clarify an interesting finding 
reported recently by 
Altfeder, Voevodin, and Roy~\cite{Altfeder}. In their experiment, the energy current in high vacuum 
between the Pt/Ir tip of the scanning tunneling microscope and gold film serving as a 
substrate was found to be anomalously large, exceeding by ten orders of magnitude the current
given by the blackbody radiation theory. An interpretation of this phenomenon 
given in  Ref.~\cite{Altfeder} involves "emission" of phonons by Au surface, facilitated 
by the electric field,  and their "tunneling" through the vacuum gap to the tip. 
Here we would like to offer an alternative mechanism that seems to be capable to account 
quantitatively for the same effect.

An important feature of this experiment is that a carbon monoxide (CO) molecule was always 
present between the Pt tip and the substrate.  From the value of the tunneling conductance, 
the gap between the "last tip atom" and the surface of the Au film
was inferred to be close to $3{\rm \AA}$ \cite {A1}.  The diameter $a$ of a CO molecule is close 
to 3.7 ${\rm \AA}$ \cite {Hirschfelder}. Thus, without the CO molecule, the distance between 
the tip and the substrate would be  about 7 ${\rm \AA}$. There is also a strong electric field 
in the gap, which is due to the work function  difference between Pt and Au  ($\approx$ 0.7 eV). 
The implication of this is that   the CO molecule must be  strongly coupled to both the tip 
and the Au surface. If we consider that this molecule serves as a mediator between the two 
thermal reservoirs (the tip and the substrate, similar to the arrangement in Fig. 1), 
the energy current between them can be 
estimated and compared with the experimental data. The coupling strength between the 
mediator and the thermal reservoirs determines 
how strongly damped the mediator is.

There are two vibration modes  associated with CO molecule attached 
to the Pt tip~\cite{Bylholder}. One of them is  the Pt-C stretching vibration with 
frequency 480 cm$^{-1}$ . The other mode is the C-O stretching motion with the 
frequency approximately 2000 cm$^{-1}$. Using the conversion coefficient 1 K = 0.7 cm$^{-1}$, 
we find that the lowest frequency 
480 cm$^{-1}$ corresponds to 685 K and the highest frequency 2000 cm$^{-1}$ corresponds 
to 2.8$\times 10^3$ K.  In the experiment, the  tip  was maintained at room temperature, 
while the temperature of the substrate was substantially lower. Thus the 
higher mode was clearly not activated and we should take the frequency of the mediator 
in our model $\omega_0\approx $ 480 cm$^{-1}$. 

Since the main frequency $\hbar\omega_0/k_{\rm B}\approx$ 685 K is substantially greater 
than the temperature of the bulk of the tip ($\approx $ 300 K) and the temperature 
of the substrate (in the range 90 - 210 K), the first order approximation of the 
heat current mediated by the CO molecule can be estimated using Eq. (\ref{quant})
\be
\label{CO}
J_{\rm CO} \approx \f{\pi^3 k_{\rm B}^4{\hat \gamma}^2}{30\hbar^3\omega_0^4}(T_1^4 - T_2^4).
\ee 
The expression for the energy current between the tip and the substrate due to phonon 
emission from the "hot" spot on the  surface
was used in Refs.~\cite{Altfeder,Wolfe} in the form
\be
\label{A}
J_{\rm A} \approx \f{\pi^5 k_{\rm B}^2}{60\hbar\theta_{\rm D}^2}(T_1^4 - T_2^4).
\ee
Here $\theta_{\rm D}$ is the Debye temperature of the substrate. The underlying physics that 
leads to these two expressions is very different, but the temperature dependence is the same. 
The analysis based on Eq. (\ref{A}) leads to a good description of the experimental results. 
Thus, the model of CO-mediated heat transfer will also give the same results if the 
prefactors of Eqs. (\ref{CO}) and (\ref{A}) are equal. This requires that
\be
\f{\hat \gamma }{\omega_0}=\f{\pi}{\sqrt {2}}\f{\hbar\omega_0}{k_{\rm B}\theta_{\rm D}}.
\ee
For gold $\theta_{\rm D}\approx$ 165 K, and since $\hbar\omega_0/k_{\rm B}\approx$ 685 K we get
\be
\label{ratio}
\f{\hat \gamma }{\omega_0}\approx 9.
\ee

Thus, other things being equal, the model based on CO-mediated heat exchange will give 
the same quantitative results 
as the model of phonon tunneling, provided that the coupling between the CO molecule and both 
the tip and substrate is  
strong enough to make the Pt-C vibrating  mode  overdamped according to Eq. (\ref{ratio}). 
This is likely the case, considering (as mentioned above) that the size of the molecule
is comparable to the gap 
between the tip and substrate and also taking into account the presence of the strong 
polarizing electric field.

The heat current given by Eq. (\ref{CO}) is $\sim 10^{10}$ times greater than  that determined by 
the black-body radiation, Eq.  (\ref{SB}), emitting from the area equal to the cross-section of 
the CO molecule $A = \pi a^2/4$, with $a \approx$ 3.7 ${\rm \AA}$, provided that
$\hat {\gamma} /\omega_0\sim 10$.

Our conclusion is that the anomalously large heat current between the STM tip and 
the substrate can be  understood just as well as the effect of mediation  by the CO molecule.
There are ways to modify the experiment in order to determine  which of the two mechanisms 
is responsible for the effect. One is to  carry out a similar measurement, but without  the CO molecule 
lodged in the tunneling gap. Another option is to use the tip and substrate made of the same metal,
so that there will be no  work function potential difference $\Delta\Phi$  between them.
The mechanism of phonon tunneling is based on the interaction between the electrically charged tip and its 
electrostatic image. The amount of charge  is mainly determined by the large electric field   
$E=\Delta\Phi_{\rm Au/Pt}/d$, where $d$ is the vacuum gap~\cite{Altfeder}. 
In the absence of such large field the mechanism of phonon tunneling should be greatly weakened.

\section{Minimum thermal conductivity}
\label{subsec:coherence}

The topic of minimum thermal conductivity can be first traced to Einstein's contribution (see 
Ref. \cite{Cahill} and references therein). In strongly disordered solids  the phonon coherence 
length may 
become of the order of interatomic distance and, obviously, cannot be reduced any further. 
Correspondingly,  
the thermal conductivity reaches its minimum value, at least as far as its dependence on such 
length is concerned.
 A detailed treatment of this problem is given in Refs. \cite{Allen, Wen}. 
The same phenomenon leads to a minimal electric conductivity in disordered conductors --
the so-called Mott-Ioffe-Regel limit~\cite{Ioffe, Mott} -- when the coherence length of the
charge carriers become comparable to the interatomic distance.

Here we would like to add another perspective on this matter using the results obtained above 
and making use of 
the scaling approach previously developed for electron transport~\cite{Thouless,Abrahams,Lee}. 
It should be emphasized that this section is not intended as a comprehensive treatment of this problem, 
but rather as a brief introduction to an alternative approach which may be useful in some systems. 

Consider a microscopic block whose  edges are along  the principal axes $\{x,y,z\}$ of the thermal 
conductivity tensor. 
We can choose the sides of the block such  that, on
average, the random phase acquired by phonons due to an inelastic interactions along the
way between the two opposite boundaries  is the same, of the order of $2\pi$, for all three pairs 
of the block boundaries. This  choice for the sides of the block
corresponds to the definition for the anisotropic phase coherence lengths 
$\ell_{\varphi,i}$ ($i=\{x,y,z\}$),
the distances over which phonons lose phase coherence. 

Let $K_{\varphi,i}$ be the thermal conductance of such a 
phase coherent volume (PCV), so that the energy current through this block
\be
j_{\varphi,i}=K_{\varphi,i}\delta T_i.
\ee
Here $\delta T_i$ is the temperature difference between the opposite edges of the PCV block. 
Notice that the notion of a temperature difference cannot be introduced for distances 
smaller than the phase coherence length. 
In order to express the macroscopic anisotropic thermal conductivity $\kappa_i$ in terms of 
the conductance of the PCV, consider a macroscopic block  with sizes $\{L_x,L_y,L_z\}$ 
obtained by fitting together $N^3$ phase coherent volumes,
so that $L_x/\ell_{\varphi,x}=L_y/\ell_{\varphi,y}=L_z/\ell_{\varphi,z}=N\gg 1$.

By virtue of Fourier's law, the heat current in the $x-$direction through this macroscopic volume is
\be
\label{Jx}
J_x=\kappa_x\Delta T_x \f{L_yL_z}{L_x}.
\ee
Here we assume a linear temperature variation across the block. On the other hand, the heat 
currents through the individual PCVs combine linearly, so that
\be
\label{Jx1}
J_x=K_{\varphi,x}\delta T_xN^2.
\ee
Taking into account that $\delta T_x=\Delta T_x/N$, we obtain
\be
\label{kappax}
\kappa_x=K_{\varphi,x}\f{\ell_{\varphi,x}}{\ell_{\varphi,y}\ell_{\varphi,z}}.
\ee
To put this result into a different perspective, the PCV  is the minimal size 
block for which one can introduce the notion of thermal conductivity. 

One can argue that the  thermal conductance of the PCV of phonons 
in an anisotropic medium is isotropic, similar to the electric conductance in anisotropic 
metals~\cite{Levin},
namely, $K_{\varphi,x}=K_{\varphi,y}=K_{\varphi,z}$. If  this assertion were true, 
the anisotropy is defined by the following relationship:
\be
\label{kappa_rat}
\f{\kappa_x}{\kappa_y}=\f{\ell_{\varphi,x}^2}{\ell_{\varphi,y}^2};\; \;\;
\f{\kappa_x}{\kappa_z}=\f{\ell_{\varphi,x}^2}{\ell_{\varphi,z}^2}.
\ee
This relationship is in agreement with the quasiclassical results obtained from the kinetic 
equations $\kappa_i \sim cv_i\lambda_i$, where  $c$ is the specific 
heat, $v_i$ is the anisotropic speed of sound, and 
$\lambda_i$ is the phonon mean free path. Considering that the mean free path is similar to 
the phase coherence length, so that $ \lambda_i\sim \ell_{\varphi,i}=v_i\tau_{\varphi }$, we get 
$\kappa_i \sim c\ell_{\varphi,i}^2/\tau_{\varphi }$. 
Since both  the relaxation (decoherence) time $ \tau_{\varphi }$ and the specific heat are 
scalars, the relationship (\ref{kappa_rat}) follows.  

Eq. (\ref{kappa_rat}) is most useful when applied to a system where at least one of the coherence 
lengths is temperature independent constant. It may be a highly disordered crystal in which 
the decoherence takes place over interatomic distances, or a layered structure in which the 
coherence length in one direction is fixed by the size of the layer. Further discussion of 
the consequences of Eq. (\ref{kappa_rat}) would be far outside the scope of this paper. 

Now we will return to
the case of highly disordered substances in which the coherence lengths in all directions 
are of the order of interatomic distances and do not change with temperature. 
There are numerous example of such substances  where the minimum thermal conductivity is reached 
at temperatures above 30 K  \cite{Cahill}.  In vitreous, 
silica- and germania-based glasses, the mean free path (or the phase coherence length)  approaches 
the interatomic distance at $T \gtrsim 100$ K.  For example, 
the phonon mean free path for amorphous selenium at 
$T \geq 50$ K is temperature independent and equal to $5\times 10^{-8}$ cm, 
which corresponds approximately to the interatomic distance in this substance~\cite{Zeller}. 

In the minimum thermal conductivity (MTC) regime the movements of the neighboring atoms 
are incoherent, so that there are no propagating  phonons. Instead, every atom is acted upon by the  
non-equilibrium environment and the mechanism of this interaction can be described by the Hamiltonian 
given by Eq. (\ref{Htot}) (see also Fig. 1). Thus, the conductance $K_{\varphi}$ of the PCV containing one 
atom can be well described by our model in which a single oscillator mediates the 
energy exchange between  two thermal reservoirs.
Then, by virtue of Eq. (\ref{kappax}) the thermal conductivity is given by
\be
\label{kappa}
\kappa_{\rm min}=K_{\varphi}/\ell_0,
\ee
where $\ell_0$ is a constant of the order of the interatomic distance and $K_{\varphi} = K$ 
is given by the general expression,  Eq. (\ref{conductance}),   and its limiting cases such as 
Eq. (\ref{class}).  

Let us consider the classical limit of high temperatures, Eq. (\ref{class}), when Eq. (\ref{kappa})
takes the form
\be
\label{kappa1}
\kappa_{\rm min} \approx \f{k_{\rm B}}{4\tau_{\rm p}\ell_0} = \f{k_{\rm B}\hat \gamma}{4\ell_0} .
\ee
This has to be compared with another  expression for the MTC which is  based on the atomic density $n$ and 
elastic constants \cite{Cahill,Wen} 
\be
\label{cahill}
\kappa_{\rm min} =0.4k_{\rm B}n^{2/3}(v_{\rm l}+2v_{\rm t}).
\ee
Here $v_{\rm l}$ and $v_{\rm t}$  are the longitudinal and transverse speeds of sound, respectively. 
This is the sum 
of the contributions of the three spatial degrees of freedom. Taking into account that
\be
n^{2/3}\sim \f{1}{\ell_0^2},\nonumber
\ee
$\kappa_{\rm min}$ can be estimated as
\be
\label{cahill2}
\kappa_{{\rm min}} \sim \f{k_{\rm B}}{\ell_0}\f{v_{{\rm av}}}{\ell_0}.
\ee
Hereafter we will drop the numerical prefactors. The characteristic time scale $\ell_0/v_{{\rm av}}$ is the 
time of flight of a phonon over the interatomic distance. The incoherence of the neighboring atoms means 
that the phonons lose their coherence over this time interval. This is exactly the meaning of the 
decoherence time $\tau_{\varphi }\sim \ell_0/v_{{\rm av}}$. In our model the  relaxation time 
$\tau_{\rm p}$ is 
determined by the strength of the coupling between the oscillator and the thermal baths. It is the 
relaxation time of the momentum of the oscillator. 
Thus, in the MTC regime $\tau_{\varphi }$ and $\tau_{\rm p}$ are equivalent quantities,
\be
\tau_{\rm p}\sim \tau_{\varphi } \sim \f{\ell_0}{v_{{\rm av}}}
\ee
and the expressions for the MTC given by Eqs. (\ref{kappa1}) and (\ref{cahill}) are qualitatively and 
even quantitatively similar.

Moreover, since the standard definition of  the Debye frequency is 
$\omega_{\rm D}\equiv k_{\rm B}\theta_{\rm D}/\hbar\sim  v_{{\rm av}}/\ell_0$, we see that in the MTC
regime the relaxation 
time in our model must be 
\be
\label{R1}
\tau_{\rm p}^{-1}=\hat \gamma \sim \omega_{\rm D}. 
\ee
The frequency $\omega_0$ is the highest frequency associated with atomic vibration and it must be also 
of the order of Debye frequency, $\omega_0\sim v_{{\rm av}}/\ell_0 \sim \omega_{\rm D}$. 
The last parameter of the model is the cut-off frequency $D$, which defines the maximum 
frequency of the modes of the thermal baths that are coupled to the mediator. 
In the context of solid substances this cut-off also must be of the order of the Debye frequency.
Thus, we come to conclusion that the range of parameters within which the Hamiltonian model given by 
Eq. (1)
is applicable to the description of the 
minimum conductivity regime  is rather narrow and is given by
\be
\label{range}
\omega_0\sim \hat \gamma \sim D\sim \omega_{\rm D}.
\ee

As an example, we can take 
the data from Ref.~\cite{Zeller}  for selenium at temperatures above 
100 K. The value of thermal conductivity  $\kappa_{\rm min}\sim 0.5\times 10^{-2}$  WK$^{-1}$cm$^{-1}$. 
Comparing this value with Eq. (\ref{kappa1}),
and taking into account that the  characteristic interatomic distance 
$\ell_0 \approx 5\times 10^{-8}$ cm, we find $\hat \gamma  \sim 10^{13}$ s$^{-1}$. 
The Debye temperature for selenium is $\theta_{\rm D}\sim 250$ K, so that 
$\omega_{\rm D}\sim 3\times 10^{13}$ s$^{-1}$
and the condition given by Eq. (\ref{R1}) is satisfied.
Thus, in a highly disordered substance the oscillator enclosed inside  the PCV
is rather overdamped,
$\omega_0\tau_{\rm p}\gtrsim 1$. Although this is only an estimate, we use it in order to illustrate the 
potential applications of our
model. A more detailed comparison with the experimental data needs to involve the specific heat 
also calculated within the framework of the same model.

\section{Josephson junctions}
\label{subsec:JJ}

Finally, we can mention that for some potential applications of our model, 
such as the Josephson junctions, all model's parameters are already
experimentally known. This enables making valuable predictions about the physical behavior 
of the corresponding system. 
In the case of the Josephson junction, the particle's coordinate $x$ in the Langevin
equation is substituted by $\phi$ which
is the phase difference 
between the wave functions describing the state of condensate of Cooper pairs 
in the contacting superconductors kept at different temperatures. Here
$\omega_0$ is
the plasma frequency and $\hat \gamma = \gamma/m = 1/RC$ with $R$ and $C$ are the junction 
resistance and capacitance, respectively~\cite{Likharev}. Characteristic values 
for $\omega_0$ can vary
between $10^{10}$ s$^{-1}$ and $10^{14}$ s$^{-1}$, depending on the current density.
 For tunnel junctions,
when $\hat \gamma/\omega_0$ is usually in the range 0.001 - 0.1, the damping is weak.
For junctions with non-tunneling conductivity and in the form of point contacts
or thin-film microbridges~\cite{Likharev_2}, $\hat \gamma/\omega_0 >> 1$ and the
damping is large.

\section{SUMMARY}

\label{conclusions}

In conclusion, we have considered the heat transport between two thermal reservoirs 
mediated by a quantum particle using the generalized quantum Langevin equation. 
Both thermal reservoirs are described as ensembles of  harmonic modes 
using the Drude-Ullersma model 
and the mediator is also treated  in the harmonic approximation.
The expressions obtained for the heat current and thermal conductance are valid for arbitrary 
coupling strength between the mediator and the reservoirs. 
The cutoff frequency, which characterizes the thermal reservoirs, can also be arbitrary.
The obtained results are analyzed for different 
temperatures regimes and different strengths of the coupling parameter. 
The dependence of the thermal conductance on the coupling strength shows a maximum
and the temperature dependence of this quantity reveals
a plateau at intermediate temperatures, similar to the classical plateau that
corresponds to the high-temperature limit. 

The results are applied to a model of a chain made out of macroscopically 
large, but finite 
subsystems, each described by the Drude-Ullersma model. 
These subsystems are coupled
to each other through a quantum mediator. As long as the subsystems are large enough, so that 
their energy changes slowly
in comparison with the relaxation rate of the mediator's energy, Fourier's law follows as a 
differential form of the energy
continuity equation. It is important to notice that at no point this derivation  relies on any 
assumptions outside the framework of the Drude-Ullersma model. Thus, it may be considered as one 
of the few 
examples of rigorous derivation of Fourier's law from the first principles, at least on the time scale
that leaves the modes of the thermal baths in thermal equilibrium.

We have applied our results to explain the observed anomalously large heat flux between STM tip and 
substrate.
Our conclusion is that the effect is due to the mediating role of the CO molecule placed in the tip-substrate gap.
We also outlined the approach by which our model can be applied in order to understand 
thermal conductivity of highly disordered substances -- the minimum thermal conductivity --
and to the non-equilibrium Josephson junction.

\section*{ACKNOWLEDGMENTS}
The authors wish to acknowledge support from the Air Force 
Office of Scientific Research. One of the authors (G.Y.P.)  is supported 
by the National Research Council Senior Associateship 
Award at the Air Force Research Laboratory.

\section*{APPENDIX}
\appendix
\setcounter{section}{1}

In the second way, at $t < 0$, the dynamical variables $x_{2i}(t)$ and $p_{2 i}(t)$ 
of the second 
bath are  determined by relations to (\ref{1st_way_x}) and (\ref{1st_way_p}), 
while $x_{1i}(t)$ and $x_{1 i}(t)$, which now incorporate the quantum system, 
are determined as
\begin{eqnarray}
\label{2st_way_x}
x_{1 i}(t) = \sum_{k=0}\sqrt{\f{\hbar}{2m_{\nu i}\nu_{1 k}}}e_i^k
(a_{1 k}^+e^{i\nu_{1 k}t} + a_{1 k}e^{-i\nu_{1k}t}) \,\,\,\,\,\,\,\,\,\,\,\,\,\,
\end{eqnarray}
and $p_{1 i}(t) = m_{1 i}\dot{x}_{1 i}(t)$. Here $e_i^k$ are orthonormal 
eigenvectors~\cite{Nieuwenhuizen}:
\begin{eqnarray}
\label{e_ik}
e_i^k = \sqrt{\f{D^2 + \nu_k^2}{D^2 + \omega_i^2}}
\f{2\Delta \omega_i \sin \phi (\omega_k)}{\pi (\omega_i^2 - \nu_k^2)} .
\end{eqnarray}

Taking into account (\ref{1st_way_T}) and (\ref{2st_way_x}), one finds
\begin{eqnarray}
\label{aver_x}
\langle x_{1i}(0)x_{1j}(0)\rangle = 
\f{\hbar}{2{\sqrt {m_{1i}m_{1j}}}}\sum_k\f{e_i^ke_j^k}{\nu_k}\coth \f{\beta_1\hbar \nu_k}{2},
\,\,\,\,\,\,\,\,\,\,
\end{eqnarray} 
\begin{eqnarray}
\label{aver_p}
\langle p_{1i}(0)p_{1j}(0)\rangle = 
\f{\hbar\sqrt {m_{1i}m_{1j}}}{2}\sum_k\nu_k e_i^ke_j^k\coth \f{\beta_1\hbar \nu_k}{2},
\,\,\,\,\,\,\,\,\,\,
\end{eqnarray} 
and $\langle p_{1i}(0)x_{1j}(0) + x_{1j}(0)p_{1i}(0)\rangle = 0$ as before. 
Using these relations, the first two sums in (\ref{Jth_less_1}) can be 
recast into
\begin{eqnarray}
\label{Jth_less_a1}
\f{1}{2m}\sum_{i=1}\f{C_{1i}}{m_{1i}}\cos (\omega_it)j_1^{(a)} = 
\f{4\hbar \gamma_1 D^2}{\pi^3m}\sum_k\Delta\nu_k(D^2+\nu_k^2) \nonumber \\
\times \sin^2\phi_k\coth\f{\beta_1\hbar\nu_k}{2} \mathcal{F}_1^{(a)}(\nu_k)
\mathcal{F}_2^{(a)}(\nu_k) \,\,\,\,\,\,\,\,\,\,\,\,\,\,\,\,
\end{eqnarray}
and
\begin{eqnarray}
\label{Jth_less_b1}
\f{1}{2m}\sum_{i=1}C_{1i}\omega_i\sin (\omega_it)j_1^{(b)} = 
\f{4\hbar \gamma_1 D^2}{\pi^3m}\sum_k\Delta\f{D^2+\nu_k^2}{\nu_k}\nonumber \\
\times \sin^2\phi_k\coth\f{\beta_1\hbar\nu_k}{2} \mathcal{F}_1^{(b)}(\nu_k)
\mathcal{F}_2^{(b)}(\nu_k) . \,\,\,\,\,\,\,\,\,\,\,\,\,\,
\end{eqnarray}
In (\ref{Jth_less_a1}) and (\ref{Jth_less_b1}), 
\begin{eqnarray}
\label{F1a}
\mathcal{F}_1^{(a)} = 
\sum_{i=1}\f{\Delta\omega_i^2\cos \omega_it}{(D^2+\omega_i^2)(\omega_i^2 - \nu^2)},
\end{eqnarray}
\begin{eqnarray}
\label{F2a}
\mathcal{F}_2^{(a)} = \sum_{n=1,2,3;\, i=1}
\f{g_n\Delta\omega_i[\mu_n\sin\omega_i t - \omega_i\cos\omega_i t]}
{(\mu_n^2+\omega_i^2)(D^2+\omega_i^2)(\omega_i^2 - \nu^2)},\,\,\,\,\,\,\,\,
\end{eqnarray}
\begin{eqnarray}
\label{F12b}
\mathcal{F}_1^{(b)} = -\p_t \mathcal{F}_1^{(a)}, \,\,\,\,
\mathcal{F}_2^{(b)} = -\p_t \mathcal{F}_2^{(a)}. 
\end{eqnarray}
Using~\cite{Prudnikov}, the above summations can be carried out accurately and the result is
\begin{eqnarray}
\label{F1a_f}
\mathcal{F}_1^{(a)}(\nu ) = \f{\pi}{2}\f{\nu}{D^2+\nu^2}\f{\cos (\nu t+\phi (\nu))}{\sin\phi (\nu)},
\end{eqnarray}
\begin{eqnarray}
\label{F2a_f}
\mathcal{F}_2^{(a)}(\nu ) = \f{\pi}{2}\f{1}{D^2+\nu^2}\sum_{n=1}^3\f{f_n}{\mu^2+\nu^2}\times \nonumber \\
\left [\mu_n\f{\sin (\nu t+\phi (\nu))}{\sin\phi (\nu)} -\nu\f{\cos (\nu t+\phi (\nu))}{\sin\phi (\nu)}\right ],
\end{eqnarray}\\
and $\mathcal{F}_1^{(b)}(\nu )$, $\mathcal{F}_2^{(b)}(\nu )$ are determined 
from (\ref{F12b}).
In the derived expressions we disregarded all contributions that are 
exponentially decaying in time. As in the first way, after substituting $\mathcal{F}_{1,2}^{(a,b)}$ 
into (\ref{Jth_less_a1}) and (\ref{Jth_less_b1}) contributions that contain 
the product $\sin (\nu t+\phi (\nu))\cos (\nu t+\phi (\nu))$ cancel each other, 
and the other time-dependent terms will be proportional to 
$\sin^2 (\nu t+\phi (\nu)) + \cos^2 (\nu t+\phi (\nu)) = 1$. As is also clear, 
the coefficient in the product $\mathcal{F}_1^{(a,b)}\mathcal{F}_2^{(a,b)}$ is inverse proportional 
to $\sin^2\phi_k$ and is canceled by similar factors in (\ref{Jth_less_a1}) 
and (\ref{Jth_less_b1}). This eliminates the dependence on initial conditions related
to whether the central particle was initially connected or not to the first bath.
These observations prove  explicitly the existence of
the steady-state in the presented mode and its uniqueness. 
Finally, replacing the summation over $k$ by the integral results in the same 
expression (\ref{P_ab_final}) for $\langle {\mathcal P}_{\nu}\rangle^{(ab)}$.


\begin{thebibliography}{99}        

\bibitem{Dhar} A. Dhar, Advances in Physics {\bf 57}, 457 (2008).
\bibitem{Dubi_Di_Ventra} Y. Dubi and M. Di Ventra, Review of Modern Physics {\bf 83},
131 (2011).
\bibitem{Jortner} {\it Molecular Electronics}, edited by J. Jortner and M. Ratner 
(Blackwell Science, Oxford, 1997).
\bibitem{Hanggi} P. Hanggi, M. Ratner, and S. Yalikari, Chem. Phys. {\bf 281}, 
111 (2002).
\bibitem{Caldeira} A.O. Caldeira and A.J. Leggett, Physica {\bf 121 A}, 587 (1983).
\bibitem{Allahverdian} A.E. Allahverdyan and Th. M. Nieuwenhuizen, 
Phys. Rev. Lett. {\bf 85}, 1799 (2000).
\bibitem{Nieuwenhuizen} Th. M. Nieuwenhuizen and A. E. Allahverdyan, 
Phys. Rev. E {\bf 66}, 036102 (2002).
\bibitem{Michel1} M. Michel, G. Mahler, and J. Gemmer, Phys. Rev. Lett. 95, 180602 (2005).
\bibitem{dubi_diventra} Y. Dubi and M. Di Ventra, Phys. Rev. E {\bf 79}, 042101 (2009).
\bibitem{Michel2} M. Michel, M. Hartmann,  J. Gemmer, and G. Mahler, Eur. Phys. J. B {\bf 34}, 325 (2003).
\bibitem{Segal_Nitzan_Hanggi} D. Segal, A. Nitzan, and P. Hanggi, J. Chem. Phys
{\bf 119}, 6840 (2003).
\bibitem{Michel3} M. Michel,  J. Gemmer, and G. Mahler, Int. J. Mod. Phys. B {\bf 20}, 4855 (2006). 
and references therein.
\bibitem{Saito2} K. Saito, Europhys. Lett. {\bf 61}, 34 (2003). 
\bibitem{Boneto} F. Bonetto, J. L. Lebowitz, and L. Rey-Bellet, 
{\it Fourier law: A challenge to theorists}, in Mathematical Physics 2000, 
edited by A. Focas, A. Grigoryan, T. Kibble, and B. Zagarlinski (Imperial College Press, London, 2000).
\bibitem{Senitzky} I.R. Senitzky, Phys. Rev. {\bf 119}, 670 (1960).
\bibitem{Mori} H. Mori, Prog. Theor. Phys. {\bf 33}, 423 (1965).
\bibitem{Ford} G.W. Ford, M. Cac, and P. Mazur, J. Math. Phys. {\bf 6}, 504
(1965).
\bibitem{Haken} H. Haken, Rev. Mod. Phys. {\bf 47}, 67 (1975). 
\bibitem{Klimontovich} Y.L. Klimontovich, {\it Statistical Theory of Open Systems} 
(Kluwer, Amsterdam, 1997).
\bibitem{Zurcher} U. Z\"{u}rcher and P. Talkner, Phys. Rev. A {\bf 42}, 3278 (1990).
\bibitem{Saito} K. Saito, S. Takesue, and S. Miyashita, Phys. Rev. E {\bf 61}, 
2397 (2000).
\bibitem{Dhar_Shastry} A. Dhar and B.S. Shastry, Phys. Rev. B {\bf 67},
195405 (2003).
\bibitem{Prosen} T. Prosen, New Journal of Physics {\bf 10}, 043026 (2008).
\bibitem{Prosen1} T. Prosen and B. \u{Z}unkovi\u{c}, New Journal of Physics {\bf 12}, 025016 (2010). 
\bibitem{Altfeder} I. Altfeder, A.A. Voevodin, and A.K. Roy, Phys. Rev. Lett. {\bf 105}, 166101 (2010).
\bibitem{Ford_Lewis_Connell} G.W. Ford, J.T. Lewis, and R.F. O'Connell, 
Phys. Rev. A {\bf 37} 4419 (1988).
\bibitem{Ullersma} P. Ullersma, Physica (Utrecht) {\bf 32}, 27 (1966); 
{\bf 32}, 56 (1966); {\bf 32}, 74 (1966); {\bf 32}, 90 (1966).
\bibitem{Weiss} U. Weiss, {\it Quantum Dissipative Systems} ( World Scientific,
Singapore, 1993).
\bibitem{Laplace} R.V. Churchill, {\it Operational mathematics}, 2nd ed.  
(McGraw-Hill, New York, 1972). 
\bibitem{Casher} A. Casher and J.L. Lebowitz, J. Math. Phys. {\bf 12}, 1701 (1971).
\bibitem{Rubin} R.J. Rubin and W.L. Greer, J. Math. Phys. {\bf 12}, 1686 (1971).
\bibitem{Benguria} R. Benguria and M. Kac, Phys. Rev. Lett. {\bf 46}, 1 (1981).
\bibitem{Dhar_Wagh} A. Dhar and K. Wagh, Eur. Phys. Lett. {\bf 79}, 60003 (2007).
\bibitem{Dhar_Sen} A. Dhar and D. Sen, Phys. Rev. B {\bf 73}, 085119 (2006).
\bibitem{Dammak} H. Dammak, Y. Chalopin, M. Laroche, M. Hayoun, and J.-J. Greffet,
Phys. Rev. Lett. {\bf 103}, 190601 (2009).
\bibitem{Landau} L.D. Landau and E.M. Lifshitz, {\it Statistical Physics}, Part 1 
(Pergamon Press, London, 1980).
\bibitem{Lifshitz} E. M. Lifshitz and L. P. Pitaevskii, {\it Physical Kinetics} (Pergamon Press, 
London, 2002).
\bibitem{Bolsterli} M. Bolsterli, M. Rich, and W. M. Visscher, Phys. Rev. A {\bf 1}, 1086 (1970).
\bibitem{Rich} M. Rich, and W. M. Visscher, Phys. Rev. B {\bf 11}, 2164 (1975).
\bibitem{Malay} M. Bandyopadhyay and D. Segal,  Phys. Rev.  E {\bf  84}, 011151 (2011).
\bibitem{A1} We thank I. Altfeder for clarifying to us some of the details of the experiment.
\bibitem{Hirschfelder} Hirschfelder, Curtiss and Bird, {\it Molecular Theory of Gases and Liquids} 
(Wiley, New York, 1954).
\bibitem{Bylholder} G. Bylholder and R. Sheets, J. Phys. Chem. {\bf 74}, 4335 
(1970).
\bibitem{Wolfe} J. P. Wolfe, {\it Imaging Phonons:Acoustic Wave Propagation in Solids}  
(Cambridge University Press, New York, 1998).
\bibitem{Cahill} D.G. Cahill, S.K. Watson, and R.O. Pohl, Phys. Rev. B {\bf 46}, 6131 (1992).
\bibitem{Allen} P. B. Allen and J. L. Feldman, Phys. Rev. B {\bf 48}, 12581 (1993).
\bibitem{Wen} W.P. Hsieh, M.D. Losego, P.V. Braun, S. Shenogin, P. Keblinski, and D.G. Cahill, 
Phys. Rev. B {\bf 83}, 174205 (2011).
\bibitem{Ioffe} A.F. Ioffe and A.R. Regel, Prog. Semicond. {\bf 4}, 237 (1960).
\bibitem{Mott} N.F. Mott, Philos. Mag. {\bf 26}, 1015 (1972).
\bibitem{Thouless} D.J. Thouless, Phys. Rep. {\bf 13C}, 93 (1974).
\bibitem{Abrahams} E. Abrahams, P.W. Anderson, D. C. Licciardello,
and T.V. Ramakrishnan, Phys. Rev. Lett. {\bf 42}, 673 (1979).
\bibitem{Lee} P. A. Lee and T.V. Ramakrishnan, Rev. Mod. Phys. {\bf 57}, 287 (1985).
\bibitem{Levin} G. A. Levin, Phys. Rev. B {\bf 70}, 064515 (2004).
\bibitem{Zeller} R.C. Zeller and R.O. Pohl, Phys. Rev. B {\bf 4}, 2029 (1970).
\bibitem{Likharev} K.K. Likharev, Sov. Phys. Usp. {\bf 26}(1), 87 (1983).
\bibitem{Likharev_2} K.K. Likharev, Rev. Mod. Phys. {\bf 51}(1), 101 (1979).
\bibitem{Prudnikov} A.P. Prudnikov, Y.A. Brychkov, and O.I. Marichev, 
{\it Integrals and Series, Vol. 1: Elementary Functions} 
(Gordon and Breach, New York, 1986).

\end{thebibliography}
\end{document}